%% file: ms.tex
\documentclass{vldb}

\usepackage{graphicx}
\usepackage{subfigure,wrapfig}
\usepackage{balance}  
\usepackage{verbatim} 
\usepackage{url}

\usepackage{indentfirst} 
\usepackage{amsmath}
\usepackage{bm}
\usepackage{times}
\usepackage{algorithm}
\usepackage{algorithmic}
\usepackage{verbatim}
\usepackage{color}
\usepackage{times}
\usepackage{array}
\usepackage{float}

\usepackage{enumitem}

\usepackage{amssymb,amsmath,amsfonts,bbm}
\usepackage{amsthm} 

\usepackage{xspace}
\usepackage[table]{xcolor}
\usepackage{multirow}

\theoremstyle{plain}
\theoremstyle{definition}
\newtheorem{definition}{Definition}[section] 

\newtheorem{proposition}{Proposition}[section]

\newtheorem{problem}{Problem}[section]

\newcommand{\skipzerolist}
    {\begin{list}{\hfil}
        {\topsep 0pt plus 1pt
           \parsep 1pt plus 1pt
           \partopsep 0pt plus 1pt
           \itemsep 0pt plus1pt
           \itemindent -0.25in }
    }

\newcommand{\skipzeroitemize}
    {
      \begin{list}{{$\bullet$}} 
        {
          \topsep 0pt plus 0pt
           \parsep 0pt plus 1pt
           \partopsep 0pt plus 1pt
           \itemsep 0pt plus 1pt}
    }

\newcommand{\minip}[1]{\vspace{0.05in} \noindent \textbf{#1}}

\newboolean{show-tr}
\setboolean{show-tr}{false}
\newcommand{\techreport}[2]{\ifthenelse{\boolean{show-tr}}{{#1}}{{#2}}}

\newboolean{showcomments}
\setboolean{showcomments}{true}
\newcommand{\markup}[1]{\ifthenelse{\boolean{showcomments}}{{\color{purple}[#1]}}{{\cut{#1}}}}
\newcommand{\cut}[1]{}
\newcommand{\todo}[1]{{\color{blue}[\textit{#1}]}}
\newcommand{\remove}[1]{}

\newcommand{\pfs}{PF-$S$} 
\newcommand{\sdnn}{PF-$AS$} 
\newcommand{\pdnn}{PF-$AP$} 
\newcommand{\wun}{\textsc{PF-Wun}}

\vldbTitle{Boosting Cloud Data Analytics using Multi-Objective Optimization}
\vldbAuthors{Fei Song, et al.}
\vldbDOI{https://doi.org/10.14778/xxxxxxx.xxxxxxx}
\vldbVolume{12}
\vldbNumber{xxx}
\vldbYear{2019}

\begin{document}


\title{Boosting Cloud Data Analytics using Multi-Objective Optimization}

\author{
\alignauthor
\hspace{-0.2in}\large{Fei Song$^\dagger$, Khaled Zaouk$^\dagger$, Chenghao Lyu$^\ddagger$, Arnab Sinha$^\dagger$, Qi Fan$^\dagger$, Yanlei Diao$^{\dagger \ddagger}$, Prashant Shenoy $^\ddagger$} \\ 
       \affaddr{$^\dagger$ Ecole Polytechnique; $^\ddagger$ University of Massachusetts, Amherst}  \\ 
    \email{ \normalsize{  $^\dagger$\{fei.song, khaled.zaouk, arnab.sinha, qi.fan, yanlei.diao\}@polytechnique.edu; $^\ddagger$\{chenghao, shenoy\}@cs.umass.edu} }  
}

\maketitle

\begin{abstract}
Data analytics in the cloud has become an integral part of enterprise businesses. Big data analytics systems, however,  still lack the ability to take user performance goals and budgetary constraints for a task, collectively referred to as task objectives, and automatically configure an analytic job  to achieve these objectives.  This paper presents a data analytics optimizer  that  can automatically  determine a cluster configuration with a suitable number of cores as well as other system parameters that best meet the task objectives. At a core of our work  is a principled  {\em multi-objective optimization} (MOO) approach that computes a Pareto optimal set of job configurations to reveal tradeoffs between different user objectives, recommends a new job configuration that best explores such tradeoffs, and employs novel optimizations to enable such recommendations within a few seconds.  We present efficient incremental algorithms based on the notion of a Progressive Frontier for realizing our MOO approach and implement them into a Spark-based prototype.
Detailed experiments using  benchmark workloads show that our MOO techniques provide a 2-50x speedup over
existing MOO methods, while offering good coverage of the Pareto frontier. When compared to  Ottertune, a state-of-the-art performance tuning system, our approach recommends configurations that yield 26\%-49\% reduction of running time of the TPCx-BB benchmark while  adapting to different application preferences on multiple objectives.

\end{abstract}

\input{intro}
\input{problem}
\input{optimization_vldb}

\input{algorithms}

\input{recommendation}

\input{experiments}
\input{moo_experiments}

\input{related_work}
\input{conclusions}


\clearpage
{
\bibliographystyle{abbrv}
\bibliography{refs/bigdata,refs/boduo,refs/db,refs/stream,refs/yanlei-diao,refs/optimization,refs/deeplearning,refs/ml-systems}}


\end{document}

%% file: intro.tex
\section{Introduction}
\label{sec:intro}
As the volume of data generated by enterprises has continued to grow, big data analytics
has become commonplace for obtaining business insights from this voluminous  data. Today, such big data analytics tasks often run  on the enterprise's private cloud or on machines
leased by the enterprise in the public cloud. Despite its wide adoption, current
big data analytics systems remain best effort in nature and typically lack the ability 
to take user objectives such as performance goals or cost constraints into account. 

Determining an optimal hardware and software configuration for a big-data analytic task based on user-specified objectives is a complex task and one that is largely performed manually. Consider an enterprise user who wishes to run a  mix of analytic tasks on their private or the public cloud. First, the user needs to choose the server hardware configuration from the set 
of available choices. Amazon's EC2 public cloud platforms currently offer over 190  hardware configurations \cite{EC2}, while Microsoft Azure offers over 30 different hardware configurations \cite{Azure}. 
These configurations differ in the number of cores, RAM size, and special hardware available.
After determining the hardware configuration, the  user also needs to determine the  software  configuration for the task by choosing various runtime parameters. For the popular  Spark runtime engine, for example, these runtime parameters include 
\textit{parallelism} (for reduce-style transformations),
\textit{number of executors}, 
\textit{number cores per executor}, 
\textit{memory per executor}, 
\textit{Rdd compression} (boolean), 
\textit{Memory fraction} (of heap space), 
to name a few.
 
The choice of a configuration is
further complicated by the need to optimize  {\em multiple}, possibly conflicting, user objectives. Consider the following real-world use cases at several large data analytics companies and cloud providers (anonymized for confidentiality) that elaborate  on these challenges and motivate our work:

{\em Use Case 1 (Data-driven Business Users)}. A data-driven security company  runs  thousands of analytic tasks in the cloud every day. The engineers managing these tasks have two objectives: keep the {\em latency} low in order to quickly detect fraudulent behaviors and also reduce {\em cloud  costs} that impose substantial operational expenses on the company. For cloud analytics, task latency can always be reduced further by  allocating more hardware resources, but this comes at  the expense of higher cloud costs. 
Hence, the engineers face the challenge of deciding the cluster configuration and other runtime parameters that balance {\em latency} and {\em cost}.

{\em Use Case 2 (Severless Databases)}. Cloud providers now offer hosted databases in the form of serverless offerings (e.g., \cite{Aurora-serverless}) where the database is turned off during idle periods, dynamically turned on when new queries arrive, and scaled up or down as the  load changes over time. A  media company that uses this serverless database to run a news site sees peak loads in the morning or as news stories break, 
and a lighter load at other times.
The news application specifies the minimum and maximum  number of computing units (CUs) to service its workload across peak and off-peak periods; it prefers to minimize cost when the load is light and expects the cloud provider to dynamically scale CUs for the morning peak or breaking news. In this case, the cloud provider needs to balance between latency under different data rates and user cost, which directly depends on the number of CUs used.  To do so, the cloud provider needs automated methods to choose appropriate configurations under different workloads that address both objectives.
Overall, choosing a configuration that balances multiple conflicting  objectives is non-trivial. Studies have show that even expert engineers are often unable to choose between two cluster options for a single objective like latency~\cite{RajanKCK16}, let alone choosing between dozens of cluster options for multiple competing objectives.


In this paper, we introduce a  {\em multi-objective optimizer}  that can automate the task of determining on optimal configuration for each task based on multiple task objectives. Such an optimizer takes as input an analytic task in the form of a dataflow program (which subsumes SQL queries) and a set of  objectives, and produces as output  a job configuration with a suitable number of cores as well as other runtime system parameters that best meet the task objectives. At the core of our work is  a principled multi-objective optimization (MOO) approach that takes multiple, possibly conflicting, objectives and computes a  Pareto-optimal set  of job configurations.
A final configuration is then chosen from this Pareto-optimal set. 

We note several differences of our work from SQL optimization: 
First, our work is  complementary to SQL optimization for database workloads. For a given query,  SQL optimization chooses a query plan, 
viewed as a dataflow program, that is then mapped to cluster  resources by choosing an appropriate number of cores and memory per core, and  configured with many parameters of a distributed  engine such as  Spark. Our optimizer addresses this later step of optimization, yielding  a {\em cluster execution plan}. 

Second, MOO for SQL queries~\cite{Ganguly:1992:QOP,Kllapi:2011:SOD,Trummer:2014:ASM,TrummerK15}  examines a {\em finite} set of query plans based on relational algebra, 
and selects the Pareto optimal ones based on estimated cost of each plan. 
In contrast, our optimizer searches through a parameter space that mixes numerical and categorial parameters, with potentially an {\em infinite} set of possible  configurations, and 
finds those  that are Pareto optimal. 
To suit the property of the parameter space, our  optimizer employs a {\em numerical optimization} approach to MOO.  

Third, our MOO-based optimizer aims to support a broad set of analytic tasks that are commonly mixed in data analytics pipelines, including  SQL queries, ETL tasks based on SQL with UDFs,  and machine learning tasks for deep analysis, all in the general paradigm of dataflow programs. To do so, our optimizer leverages recent machine learning based modeling approaches~\cite{VanAken:2017:ADM,Zhang:2019:EAC} that can automatically learn a predictive model for each  objective  using  the runtime behavior of a user task (i.e., runtime metrics), without necessarily requiring the use of query plans. 
In particular, we view our MOO work as a synergy with  recent work on workload modeling. Such a synergy is reminiscent of the past work on SQL optimization: our  MOO framework is  analogous to the Dynamic Programming based optimization framework, although it has been extended to the multi-objective settings due to the needs of today's cloud analytics, while recent modeling work is analogous to cost modeling of SQL query plans but extended to automated learning of such cost models from runtime observations.  As we show, working with learned models brings new challenges for optimization. 

%

More specifically, our design of a multi-objective optimizer addresses the following technical challenges:

1. {\em Infinite Parameter Space}: There are potentially infinite configurations in our parameter space, but only a small fraction of them belong to the Pareto set---most configurations are dominated by some Pareto optimal configuration for all objectives. 
Hence, we must address the challenge of efficiently searching through an
infinite parameter space to find these Pareto optimal configurations. 

2. {\em Coverage of the Pareto Frontier}: The Pareto set over the multi-objective space is also called the {\em Pareto frontier}. Since we aim to use the Pareto frontier to recommend a new configuration that best explores tradeoffs between different objectives, the frontier should provide good coverage of the overall objective space and have a  fine resolution for the regions when the tradeoffs are significant. 
As we show later, classical MOO algorithms~\cite{marler2004survey} often fail to provide sufficient coverage of the Pareto frontier. 

3. {\em Efficiency}: 
Since our optimizer uses learned models of user objectives, it has to handle the high complexity of such models (e.g., using Deep Neural Networks) and  frequent updates of these models as new training data becomes available. Before running a user task, the learned models may have changed and the optimizer may have to recompute the Pareto frontier in order to make recommendations.   Therefore,  the speed of computing the Pareto frontier, e.g., within {\em a few seconds}, is crucial for adapting to bursty data loads quickly (in the serverless database case) or reducing the delay of starting a recurrent workload at a scheduled time. 
Most existing MOO algorithms, including Weighted Sum~\cite{marler2004survey}, Normal Constraints~\cite{marler2004survey}, and Evolutional Methods~\cite{Emmerich:2018:TMO}, are not designed to meet such stringent performance requirements.

By addressing the above challenges, our paper makes the following contributions: 

(1) We address the  infinite search space issue by presenting a new approach for incrementally transforming a MOO problem to a set of constrained optimization (CO) problems, where each CO problem can be solved individually to return a Pareto optimal point.

(2) We then address the coverage and efficiency challenges by designing Progressive Frontier (PF) algorithms to realize our approach. ($i$)~Our first PF algorithm is designed to be {\em incremental}, i.e., gradually expanding the  Pareto frontier as more computing time is invested, and {\em uncertainty-aware}, i.e., returning more points in regions of the frontier that lack sufficient information. ($ii$)~We also develop an {\em approximate}  PF algorithm that given complex learned models, solves each CO problem efficiently  based on these models.  ($iii$)~We finally devise a {\em parallel}, approximate  PF algorithm to further improve efficiency.


(3)  We implement our algorithms into a Spark-based prototype. 
Evaluation results using benchmarks for batch and streaming workloads show that our approach  produces a Pareto frontier in less than 2.5 seconds (2-50X faster that other MOO methods~\cite{marler2004survey,Emmerich:2018:TMO}), provides greater coverage over the frontier, and enables exploration of tradeoffs such as cost-latency or latency-throughput. 
When compared to Ottertune~\cite{VanAken:2017:ADM}, a state-of-the-art performance tuning system, our approach recommends configurations that yield 26\%-49\% reduction of total running time of the TPCx-BB benchmark~\cite{TPCx-BB} while adapting to different application preferences on multiple objectives and being able to accommodate a broader set of models. 
As database research continues to deliver new results on learned models~\cite{DuanTB09,Marcus:2019:PDN,Marcus:2019:NLQ,VanAken:2017:ADM,Zhang:2019:EAC}, the generality of our optimizer allows it to achieve even better results once the new learned models are made available.

%% file: problem.tex
\section{Background and Motivation}
\label{sec:problem}


In this section, we discuss requirements and constraints  from real-world use cases that motivate our system design. 

\vspace*{-0.1in}
\subsection{Requirements and Design Choices}
\label{subsec:background}

Our work targets a broad set of analytic tasks such as SQL queries with user-defined functions (UDFs), ETLs tasks based on SQL, and machine learning-based  analytics. We model these analytic tasks as {\em dataflow} programs that are designed to run on big data systems such as Spark~\cite{Zaharia+2012:rdd,Zaharia+2013:discretized} and Flink~\cite{Flink-recovery2015}.
We assume that each task has user- or provider-specified objectives, referred to as {\em task objectives}, that need to be optimized during execution. Common objectives include latency, throughput, and resource (i.e., cloud) costs. 
A task can specify multiple objectives and it is possible for these objectives to conflict.  
To execute the task, the system needs to determine a  {\em cluster execution plan} with system parameters instantiated.  As stated before, these parameters control the degree of parallelism (the number of cores or executors), memory allocated to each executor or data buffer, granularity of scheduling, compression options, shuffling strategies, etc. 
An executing task using this plan is referred to as a {\em job} and  the system parameters are collectively referred as the {\em job configuration}. 

The overall goal of our multi-objective optimizer is: {\em given a  user data flow program and a  set  of objectives, compute a job configuration that optimizes these objectives 
and adapt the configuration quickly if either the task load or task objectives change.} 


In particular, our work is designed to address the following requirements of real-world analytics tasks.

1. {\em Recurring workloads.}  It is common for analytic environments to see repeated
jobs in the form of daily or hourly batch jobs. In some cases, recurring jobs  
have dependencies that trigger other jobs upon completion, yielding a pipeline of 
repeating jobs.
Sometimes stream jobs can be repeated as well: under the lambda architecture, the batch layer runs  to provide  perfectly accurate analytical results, while the speed layer offers fast approximate analysis over live streams; the results of these two layers are combined to serve a model. As old data is periodically rolled into the batch job, the streaming job is restarted over  new data with a clean state. 
Our  optimizer is designed  
to be invoked prior to each execution of a recurring job with the goal
of choosing a configuration that improves performance towards target objectives.

\cut{
{\em  Similarities across workloads.}
Similarities across workloads arise for several reasons: 
(a)~Early processing tasks in analytics pipelines are often similar; they can be shared by tens of downstream tasks.
(b)~Most workloads are parameterized, i.e., generated from a set of templates with the parameters set to appropriate values by each user.
Such similarities  offer an opportunity for the optimizer to improve prediction accuracy, even under a blackbox approach. }

2. {\em Serverless database workloads.} 
As noted in Section \ref{sec:intro}, serverless databases (DBs) are becoming common in cloud environments.
Each invocation of a serveless DB  by the cloud platform requires a configuration that satisfies multiple objectives---to provide low query latency to  end-users while using the least cost cloud configuration for the expected load. Furthermore, auto scaling features of serverless DBs imply that new configurations need to be computed quickly to react to load changes. \cut{Note that in the serverless scenario, the task objectives are specified  by the provider rather than the end-user, since provider decides the configuration for each user and needs to do so in a cost-effective manner to maximize revenue. }
Our optimizer is  designed to support serverless DBs by quickly (re)computing a configuration at each invocation or auto-scaling to a larger cluster instance.

In both of the above scenarios, configurations need to be recommended under {\bf stringent time requirements}, e.g., within a few seconds, in order to minimize the delay of starting a recurring job, or  invoking or scaling a serverless DB.
Such time constraints distinguish our MOO work for the use by a cloud optimizer from the theoretical work on MOO~\cite{marler2004survey,Emmerich:2018:TMO} in the optimization community. 

3. {\em  Mixed workloads}. 
Besides SQL queries,  cloud analytics  today often involve large ETL jobs  for data cleaning, transformation, and integration, as well as machine learning  tasks for deep analysis. 
Even SQL queries make extensive use of user-defined functions, as reported lately~\cite{RajanKCK16}.
In order to support such diverse analytic  workloads, our work leverages recent machine learning based modeling approaches~\cite{VanAken:2017:ADM,Zhang:2019:EAC} that can automatically learn a predictive model for each task objective based on runtime observations collected from job execution. 
The implication on our design  is that the optimizer takes as input predictive models for target objectives, where each  model is learned to predict the value of a target objective based on the configuration of all  system parameters. 
Designing a MOO solution over learned models that are highly complex, e.g., using Deep Neural Networks (DNNs), and updated frequently, is challenging given our {\bf coverage and efficiency requirements}.


\subsection{Relation to Prior Work} 
\label{subsec:priorwork}
We next describe the relation of our MOO work  to the most relevant prior work.

\noindent {\bf Single-objective performance tuning.} Recent systems such as OtterTune~\cite{VanAken:2017:ADM} and CDBTune~\cite{Zhang:2019:EAC} address performance tuning of SQL queries for objectives such as minimizing latency.  They determine how to set the parameters of a RDBMS by modeling the objective as a function of the parameters and then iteratively exploring new configurations to update the model and move the observed performance toward the optimum of the objective.  These systems differ from our work in two main aspects:

1) {\em Single objective optimization}. Both OtterTune and CDBTune  are inherently designed for 
optimizing a single, fixed objective, while we seek to optimize multiple objectives.
OtterTune can consider only  one objective (e.g., latency) for optimization~\cite{VanAken:2017:ADM}.  CDBTune considers both latency and throughput but uses a fixed, weighted approach to compose a single objective from them, e.g., using weights \cite{Zhang:2019:EAC}.
As we will show, using a weighted approach to reduce a MOO problem to a fixed, single objective (SO) optimization  problem entails significant loss of  information regarding tradeoffs between different objectives and yields suboptimal solutions. 

2) {\em Integrated modeling and optimization:}
OtterTune and CDBTune integrate modeling and optimization steps into a single, long tuning session for each query workload. Such a session couples modeling and optimization via iterative exploration of new configurations and takes 15-45 min to run [37,43]. In contrast, our target applications impose stringent time constraints and require new configurations to be computed in seconds rather than tens of minutes. 

To meet these stringent time constraints, we argue for decoupling  modeling and optimization into two separate steps.  We assume that the time-consuming modeling step
is performed asynchronously in the background whenever  new training data becomes available. 
The MOO step runs separately on-demand and uses the most recent models to compute a new configuration with the delay of a few seconds. Such decoupling allows fast computation of new configurations with latencies that are not possible in approaches that integrate the two steps. The effectiveness of such decoupling was recently shown in a preliminary demo of our system~\cite{UDAO-vldb-demo}.

The implications of this  change are two-fold: ($i$) It enables the MOO to be general and not tied a specific modeling approach or tool.
This frees the modeling engine to use any appropriate modeling tool, and any future improvements in modeling automatically improve the efficacy of the MOO as well. Integrated approaches do not allow this flexibility---OtterTune's optimization method is tied to the Gaussian Process (GP) modeling, while CDBTune's optimization method is tied to its Reinforcement Learning (RL) model. 
As we shall show  in this paper, our optimization solution works as long as the learned models can be represented as a regression function on system parameters, e.g., using a GP or a DNN.  ($ii$) The  modeling engine can train a new model in the background as new training data becomes available. When the MOO solution needs to be run for a given task, the model for this task may have been updated, and hence the speed to compute a Pareto frontier based on the new model is a key performance goal.



\noindent {\bf Learning-based  modeling.} 
Our MOO is designed to work with any applicable modeling approach that can produce predictive models for task objectives.  Since our focus is on MOO rather than modeling, we leverage recent work on automatic model learning~\cite{DuanTB09,VanAken:2017:ADM,Zhang:2019:EAC} for our MOO. 
Such modeling work makes two assumptions:
(1) A model for each objective can be learned from runtime observations, e.g., latency as the response to different configurations of system parameters.  (2) Models for different objectives can be built as regression functions over the full set of system parameters. Today's analytics engines have large numbers (10s or 100s) of parameters, as reported for DBMSs~\cite{VanAken:2017:ADM,Zhang:2019:EAC} and as we observed for the Spark engine. These parameters  may contribute differently to the objectives,  which are captured by different weights in the learned models. Further, for a specific objective unimportant parameters  can be filtered via feature selection~\cite{VanAken:2017:ADM}.  

Among modeling work, OtterTune~\cite{VanAken:2017:ADM} best matches the needs of our modeling engine: it can  automatically learn a model for each objective as a regression function on system parameters. It is better than previous work that employs hand-crafted models~\cite{RajanKCK16,Venkataraman:2016:EEP}, hence hard to generalize. It is also better than iTuned~\cite{DuanTB09}, which uses GP models to search for optimal configurations, but fails to train models using data from the  history and hence offers inferior performance. Finally, CDBTune~\cite{Zhang:2019:EAC} employs a RL approach and hence is subject to the limitation that the reward of choosing a configuration is a scalable value. To cope with it, CDBTune uses fixed weights to combine latency and throughput into a single objective in order to build the reward function. As such, it cannot return a regression function for {\em each} objective as required for our MOO.

Therefore, our current prototype uses  OtterTune as the default approach for the modeling engine while still supporting other modeling solutions, i.e., DNNs. 
Our experiments demonstrate the benefits of our MOO solution over the optimization solution of OtterTune when both systems use the same set of learned models.

\begin{figure}[t]
	\centering
	\includegraphics[width=0.45\textwidth]{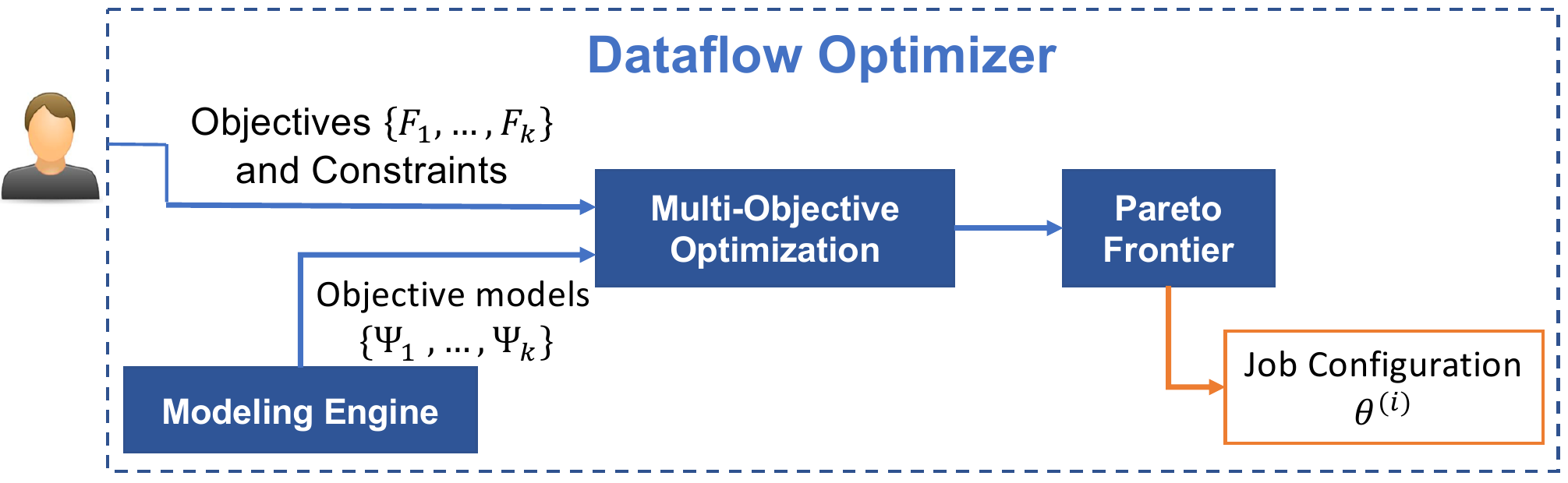}
	\vspace{-0.1in}
	\caption{\small Overview of a dataflow optimizer.}
	\vspace{-0.2in}
	\label{fig:optimizer}
\end{figure}

\subsection{System Design}
\label{subsec:system}

Figure~\ref{fig:optimizer} shows the design of our {\em dataflow optimizer} based on the above requirements. It takes as input a set of  objectives for a task,
denoted by $(F_1, \ldots, F_k)$, and optionally value constraints on these objectives,  [$F_i^L, F_i^U$]. It also takes a set of task-specific predictive models, $(\Psi_1, \ldots, \Psi_k)$,  one for each objective,  and computes a job configuration that optimizes these objectives.   

When a job runs for the first time, no predictive models will be available for the job.
The job is assumed to run with a default configuration $\bm{x}^1$.
Our optimizer assumes that  a separate modeling engine will collect traces during the job execution and then use these traces to design task-specific models.  
Two types of traces are assumed to be collected: (i) system-level metrics, e.g., from the Spark engine, such as  records read and written, bytes spilled to disk, and fetch wait time, and (ii) observed values of task objectives such as latency and compute cost. 
The modeling engine is assumed to use these traces to compute task-specific regression models $(\Psi_1, \ldots, \Psi_k)$,  one for each user objective. Our optimizer is designed to work with any modeling approach that can provide such models---our current implementation uses models learned by  OtterTune \cite{VanAken:2017:ADM} as well as our own custom DNN implementations. Regardless of the approach used, the model training is done offline and separately from the MOO path.

Given the task-specific predictive models, the multi objective optimization (MOO) module searches through the space of configurations and computes a set of Pareto-optimal configurations for the job.  
Based on insights revealed in the Pareto frontier, the optimizer
chooses a new configuration, $\bm{x}^2$, that meets all user constraints and best explores the tradeoffs among different objectives. Future invocations of the task run with this new configuration.

If the user decides to adjust the constraints on the objective (e.g., specifying new bounds [$F_i^L, F_i^U$] to increase the throughput requirement in order to adapt to higher data rates), the MOO can quickly return a new configuration from the computed Pareto frontier. As the modeling engine continues to collect additional samples from recurring executions of this task as well as others, it may periodically update the task's predictive models. Upon the next execution of the task, if updated models become available, the Pareto frontier will be recomputed by re-running the MOO and a new configuration will be chosen.

%% file: optimization_vldb.tex
\section{Progressive Frontier Approach}

\label{sec:optimizer}
In this section, we present a new Progressive Frontier framework for solving a multi-objective optimization problem, and then use this concept  to design fast algorithms in the next section. 


\subsection{Problem Statement}
\label{subsec:moo_problem}
We begin with a mathematical definition of the multi-objective optimization problem for a given user task.

{\problem{\textbf{Multi-Objective Optimization (MOO).}
\begin{eqnarray}
\label{eq:mult_obj_opt_def}
  \arg \min_{\bm{x}} &f(\bm{x})=& {\left[
        \begin{array}{l}
          F_1(\bm{x}) = \Psi_1(\bm{x}) \\ 
          ...\\
          F_k(\bm{x}) = \Psi_k(\bm{x}) 
        \end{array}
      \right]} \\
 \nonumber s.t. & &  {\begin{array}{l}
          \bm{x} \in \Sigma \subseteq \mathbb{R}^d  \\ 
          F_i^L \leq F_i(\bm{x}) \leq F_i^U, \,\,\, i=1,...,k\\
        \end{array}}
\end{eqnarray}
}}\noindent{where $\bm{x}$ is the job configuration with $d$ parameters,   
$\Sigma \subseteq \mathbb{R}^d$ denotes all possible job configurations. 
Further, $F_i(\bm{x})$ generally denotes  one of the $k$ objective functions,  while  $\Psi_i(\bm{x})$  refers particularly to the predictive model learned for this  objective. 
If a task objective favors larger values,  we add the minus sign to the objective function to transform it to a minimization problem. Optionally, there can be a number of inequality constraints on the parameter vector, $\bm{x}$. 
Note that our problem involves the the $d$-dimensional {\bf parameter space}, $\Sigma$, and the $k$-dimensional {\bf objective space}, $\Phi$, where each dimension corresponds to an objective. }

In general, multi-objective optimization (MOO) leads to a set of solutions rather than a single optimal solution. Hence, the notion of optimality in MOO settings is based on the following concepts:

\definition{\textbf{Pareto Domination:} In the objective space $ \Phi$, a point \bm{$f'$} Pareto-dominates another point \bm{$f''$} iff $\forall i \in [1, k], f_i' \leq f_i''$  and $\exists j \in [1, k], f_j' < f_j''$.

\definition{\textbf{Pareto Optimal:} In the objective space $\Phi$, a point \bm{$f^*$} is Pareto optimal iff there does not exist another point \bm{$f'$} that Pareto-dominates it.

\definition{\textbf{Pareto Set (Frontier):} For a given job, the Pareto set (frontier) $\mathcal{F}$ includes all the Pareto optimal points in the objective space $\Phi$, and is the solution to the MOO problem. \cut{We refer to  a Pareto  optimal point  as a {\em reference point}, $F^0_i$, if it achieves the minimum for the objective $f_i$.}
\vspace{0.05in}

The MOO problem  is characterized by a mapping ($\Psi_1, \cdots \Psi_k$)  from the numerical parameter space, $\Sigma$, to the numerical objective space, $\Phi$. Through this mapping, we want to find those configurations  in the parameter space  $\Sigma$ that lead to Pareto-optimal points in the objective space $\Phi$.

Besides the above definition, we impose performance requirements   to suit the needs of a multi-objective optimizer for cloud analytics. As noted in Section 1, the Pareto frontier needs to be computed with {\bf good coverage} in order to enable recommendations of best configurations,  and {\bf high efficiency}  to respond to the need of starting a new database instance or adapt to bursty data loads quickly in the serverless database case, or to reduce the delay of starting a recurrent workload at a scheduled time. 

\subsection{Existing Numerical MOO Methods}
\label{subsec:moo_survey}

\emph{Weighted Sum (WS)}~\cite{marler2004survey}: 
The WS method~\cite{marler2004survey}
transforms the MOO problem into a single-objective optimization problem by distributing the weights (preferences) among different objectives.
The WS method tries a number of weight distributions and for each computes the optimal solution. It returns the collection of solutions as the Pareto set. 
A major issue with WS  is the {\bf poor coverage} of the Pareto frontier~\cite{das1997a, messac2012from}, which is undesirable if one wants to understand tradeoffs across the entire Pareto frontier.

\cut{
It is also known that WS is not able to obtain points on non-convex portions of the Pareto Frontier \todo{in the criterion space}~\cite{das1997a, messac2012from}.
Adaptive Weighted Sum (AWS) approach, which is a variant of weighted sum approach could handle non-convex function by recursively calling weighted sum approach for the segment which does not have any solutions returned. AWS has much longer running time than WS since WS is invoked many times.
}

\emph{Normalized Constraints (NC)}~\cite{marler2004survey}: To address the coverage issue, the NC method~\cite{messac2003nc} provides a set of evenly spaced Pareto optimal points on the frontier $\mathcal{F}$. 
It divides the objective space into an evenly distributed grid and probes the grid points to have even coverage of the objective space. 
However, NC suffers from  {\bf efficiency} issues. 
First, it uses a pre-determined parameter, $N^p$, to specify how many Pareto  points  to explore, which can grow exponentially with with the dimensionality of $\Phi$. This parameter can impose high computational cost (if set too high) or produce inaccurate Pareto frontiers (if set too low).
Second, NC is not an incremental algorithm: if  the Pareto frontier built form $k$ probes does not provide sufficient information about the tradeoffs, then the next run with  $k' (> k)$ of probes will start the computation from scratch. 

\emph{Evolutionary Methods} (Evo)~\cite{Emmerich:2018:TMO}
 are randomized methods  to approximately compute the frontier set. 
As evolutionary methods are used to build an optimizer in our setting, they suffer from two issues:
First, it is not an incremental algorithm. Like the NC method, if  the Pareto frontier built from $k$ probes does not provide sufficient information about the tradeoffs, then the next run with $k' (> k)$ probes will start  from scratch. Since the optimizer does not know the sufficient number of probes in advance, it needs to start from a smaller value and try larger values if more information is needed. Hence, evolutionary methods will suffer from the {\bf efficiency} issue by not being able to reuse the previous result.   
Second, a more severe problem is {\bf inconsistency}: The Pareto frontier built from $k'$ probes can be inconsistent with that built from $k$ probes, as our experimental results show. This causes a major problem in  recommending configurations: if the optimizer uses the Pareto frontier from $k$ probes to make a recommendation due to stringent time requirements, it can be invalidated completely later as the Pareto frontier is recomputed to include more points.

\subsection{Overview of Our Approach}
\label{subsec:progressive}

To meet the aforementioned performance requirements, we introduce a new approach, called {\bf Progressive Frontier}. It incrementally transforms the MOO problem into a series of constrained single-objective optimization problems, which can be solved individually. In this section, we formally describe our approach and show its correctness.
We first show how to generate a single Pareto optimal point by solving a constrained optimization (CO) problem derived from the MOO; later on, we use this building block to incrementally generate a series of CO problems to compute a number of Pareto-optimal points on the Pareto frontier. Our approach overcomes limitations of  traditional CO by automatically choosing the constraint threshold and incrementally improving the solution set. 

{\problem \label{prob:co}{\textbf{Constrained Optimization (CO).} Given the MOO defined in Formula~\ref{eq:mult_obj_opt_def}, its constrained optimization  is a single objective optimization problem $P_{\mathcal{C}}^{(i)}$ defined as:
\begin{equation}
\begin{aligned}
\bm{x}_{\mathcal{C}} = \hspace{0.2cm} & \underset{\bm{x}}{\text{\bf arg min}}\hspace{0.3cm}F_i(\bm{x}) ,\\
 \text{\bf subject to}\hspace{0.4cm} & \mathcal{C}_j^L \leq F_j(\bm{x}) \leq \mathcal{C}_j^U, \hspace{0.1cm} j \in [1,k]
\end{aligned}
\label{eq:co}
\end{equation}
}where $\mathcal{C}_j$ is constraint on the $j$-th objective, $\mathcal{C}=\{[\mathcal{C}_j^L,  \mathcal{C}_j^U]\ | \ j \in [1,k]\}$ is the set of constraints on all the objectives.}

{\proposition \label{local_pareto}  The solution to the above CO problem is Pareto optimal within the hyperrectangle formed by the constraints, $\mathcal{C}=\{[\mathcal{C}_j^L,   \mathcal{C}_j^U]$, $j \in [1, k]$\}.}
\vspace{0.2cm}

The result is easy to prove---within the hyperrectangle, we have minimized the objective $i$ and hence, no other points can dominate this solution in the same hyperrectangle. \cut{Hence, the solution to CO is also Pareto optimal since it is Pareto optimal within the hyperrectangle constructed by the constraints $[\mathcal{C}_j^L,   \mathcal{C}_j^U]$.}


We next explain how exactly we construct the CO problem given $k$ objectives. To present our solution, we repeat some definitions used in the NC method~\cite{messac2003nc} below.

\definition{{\bf Reference Point:} \bm{$r_i$} is a special Pareto optimal point if it achieves the minimum for  objective $F_i$.}
\cut{
Since finding \bm{$r_i$} is a single objective optimization problem, it can be solved directly. For a $k$-objective optimization problem, there exist $k$ different reference points.
}
\definition{{\bf Utopia and Nadir Points:} \bm{$r^1$}, $\ldots,$  \bm{$r^k$} $\in \Phi$} are $k$ reference points. A point \bm{$f^U$} $\in \Phi$ is a Utopia  point iff for each $i = 1,2,...,k$, $f^U_i$ $=\min_{l=1}^k \{r^l_i\}$}. A point \bm{$f^N$} $\in \Phi$ is a Nadir point iff for each $i = 1,2,...,k$, $f^N_i$ $=\max_{l=1}^k \{r^l_i\}$}. 
\vspace{0.1in}

The Utopia point and Nadir point indeed are the two extreme points that form a hyperrectangle in the objective space. Every point in this hyperrectangle Pareto dominates the Nadir point and is dominated by the Utopia point. The Pareto frontier can be of arbitrary shape within this hyperrectangle. 


{\proposition \label{first_hyperrectangle} For the hyperrectangle enclosed by the Utopia and Nadir points, no Pareto optimal points outside the hyperrectangle (if they exist) can dominate any  point within it. Further, in the 2-dimensional case, for any hyperrectangle enclosed by a pair of known Pareto optimal points, the same conclusion holds.}
\vspace{0.2cm}

\noindent
Due to space constraints, we refer the reader to \techreport{Appendix~\ref{sec:proof}}{our technical report~\cite{Song2020}} for  the proof of this and other propositions.

Next we describe a method to find {\bf one} Pareto point in the space enclosed by the Utopia point and the Nadir Point. 
\cut{
For now, one can interpret uncertain space as the hyperrectangle enclosed by the Utopia point and the Nadir Point: a Pareto point can exist anywhere within it.
}
\definition{{\bf Middle Point Probe:}  Given a hyperrectangle formed by \bm{$f^U}$ = $(f_1^U, \ldots, f_i^U, \ldots, f_k^U)$ and  \bm{$f^N$} = $(f_1^N,$ $\ldots, f_i^N, \ldots, f_k^N)$, the middle point \bm{$f^M$} bounded by \bm{$f^U}$ and \bm{$f^N$} is defined as the solution to constrained optimization of:
	\begin{equation}
	\begin{aligned}
	\bm{x}_{\mathcal{C}}=  \hspace{0.2cm} & \underset{\bm{x}}{\text{\bf arg min}}\hspace{0.3cm}F_i(\bm{x}) ,\\
	\text{\bf subject to}\hspace{0.2cm} 
	& f_j^U \leq F_j(\bm{x}) \leq  \frac{(f_j^U + f_j^N)}{2}, \hspace{0.2cm} j \in [1,k]. 
	\end{aligned}
	\label{eq:mpp}
	\end{equation}}
	
\noindent {As before, we use $\mathcal{C}^{\bm{M}}$ to denote the constraints in Eq. \ref{eq:mpp}.
In theory, we can choose any $i$ to be the objective of the middle point probe.}

\begin{figure}[t]
	\vspace{-0.1in}
	\hspace{-0.2in}
	
	\begin{tabular}{cc}
		\hspace{-0.2in}
		\subfigure[\small{Middle point probe.}]
		{\label{fig:us_example_2d_1}
			\includegraphics[width=4.2cm]{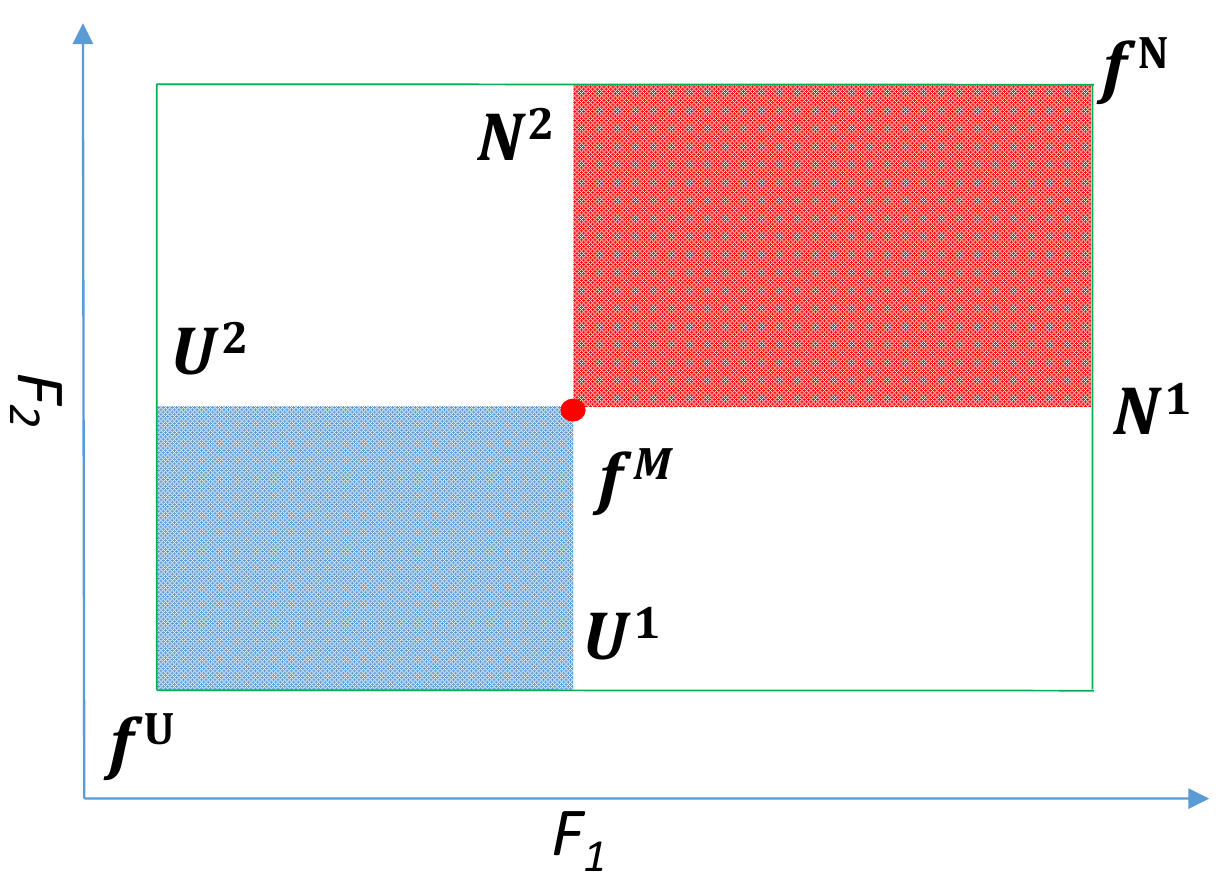}}
		
		&
		\hspace{-0.2in}
		\subfigure[\small{multiple Pareto points.}]
		{\label{fig:us_example_2d_2}\includegraphics[width=4.2cm]{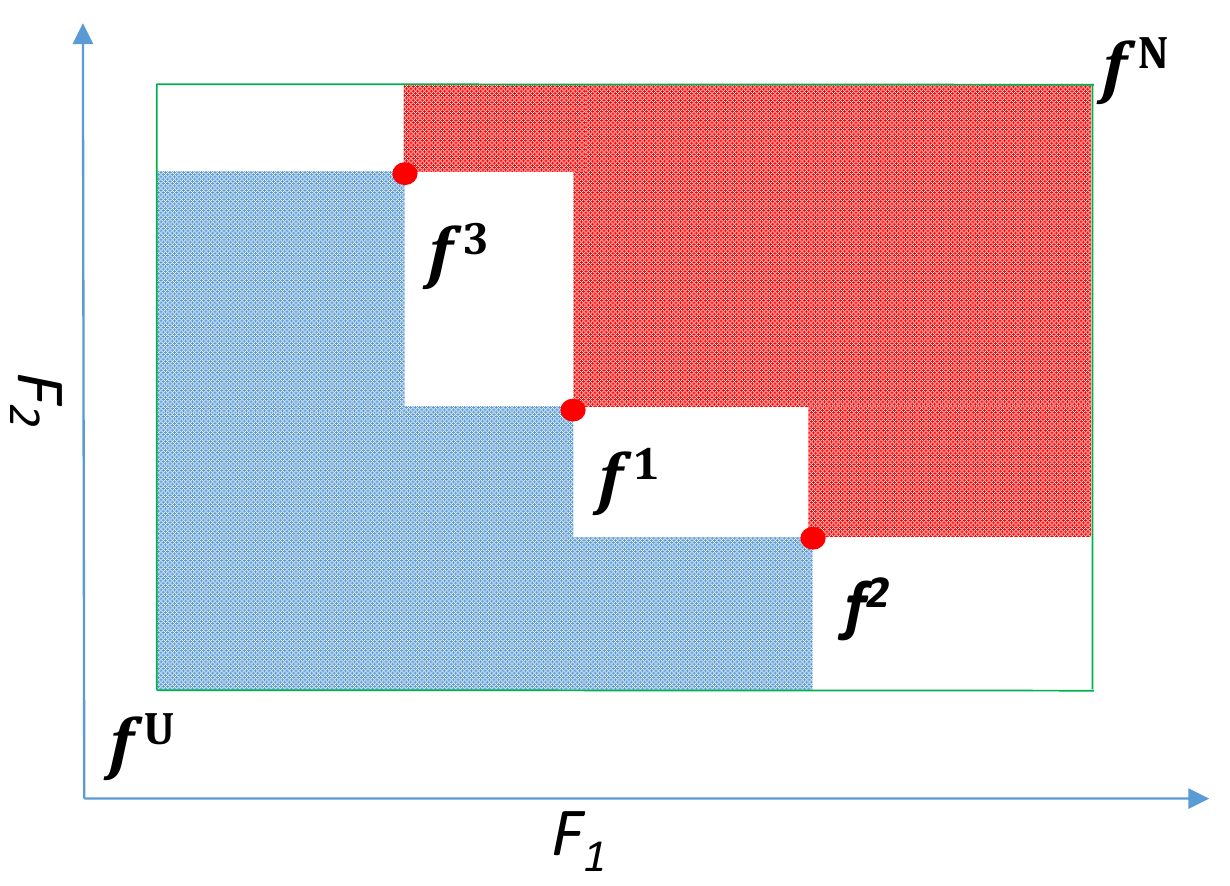}}
		
	\end{tabular}
	\vspace{-0.2in}
	\caption{\small Uncertain space in 2-dimension. \bm{$f^U$} and \bm{$f^N$} are the Utopia and Nadir points, respectively. In (a), \bm{$f^M$}  is the solution to the middle point probe. In (b), (\bm{$f^1$}, \bm{$f^2$}, \bm{$f^3$})  represent the solutions of a series of middle point probes.}
	\label{fig:us_example}
	\vspace{-0.1in}
\end{figure}

{\proposition \label{mpp_optimality} If we can not find the middle point \bm{$f^M$}, then there are no Pareto optimal points within the hyperrectangle enclosed by the constraints $\mathcal{C}^{\bm{M}}$. In the presence of \bm{$f^M$}, by Proposition \ref{local_pareto} and  Proposition \ref{first_hyperrectangle}, \bm{$f^M$} is a Pareto optimal point in the 2D case, and it is a candidate point regarding Pareto optimality, subject to further filtering, in high-dimensional cases. 
\vspace{0.2cm}

\noindent The detailed proof is given in \techreport{Appendix~\ref{sec:proof}}{our technical report~\cite{Song2020}}. Note that in high-dimensional cases, our result is similar to classical MOO algorithms like NC~\cite{marler2004survey} in that the algorithm can return only likely candidates for further pruning due to the complexity of dividing high-dimensional space. 

\cut{
In the presence of \bm{$f^M$}, the sub-hyperrectangle enclosed by \bm{$f^M$} and \bm{$f^U$} contains only the points that dominate \bm{$f^M$}, as depicted by the blue rectangle in Figure \ref{fig:us_example_2d_1}. Due to the Pareto optimality of \bm{$f^M$}, we are certain that no Pareto points exist there. Similarly the sub-hyperrectangle enclosed by \bm{$f^M$} and \bm{$f^N$}, depicted by the red rectangle in Figure \ref{fig:us_example_2d_1}, contains only points dominated by \bm{$f^M$}; hence  no Pareto points can exist there. 
}


To quantify the effect of finding \bm{$f^M$} on the objective space, we introduce the following notion: 

{\definition{\textbf{Uncertain Space:} Given a hyperrectangle formed by the Utopia Point \bm{$f^U}$ and Nadir Point \bm{$f^N}$ in the objective space, the uncertain space is defined as the space within this hyperrectangle 
that encloses all possible shapes of the Pareto frontier that are consistent with the current set of Pareto optimal points available.
\label{def:uncertainspace}}}

Intuitively, one can interpret the uncertain space as subregions of the objective space where a Pareto point may exist anywhere within it. 
In the presence of \bm{$f^M$}, the sub-hyperrectangle enclosed by \bm{$f^M$} and \bm{$f^U$} contains only the points that dominate \bm{$f^M$}, as depicted by the blue rectangle in Figure \ref{fig:us_example_2d_1}. Due to the Pareto optimality of \bm{$f^M$}, we are certain that no Pareto points exist there. Similarly the sub-hyperrectangle enclosed by \bm{$f^M$} and \bm{$f^N$}, depicted by the red rectangle in Figure \ref{fig:us_example_2d_1}, contains only points dominated by \bm{$f^M$}; hence  no Pareto points can exist there. 
Hence by  finding \bm{$f^M$}, we can reduce the uncertain space from the original hyperrectangle (enclosed by \bm{$f^U}$ and \bm{$f^N}$) by discarding these two colored sub-hyperrectangles.
\cut{
In Figure \ref{fig:us_example_2d_2}, we illustrate the uncertain space given a set of Pareto optimal points in $2$D case. 
Each of the Pareto points can divide the rectangle enclosed by (\bm{$f^U}$, \bm{$f^N}$) into four parts, and the union of the colored regions is where no Pareto optimal points exist. The complement of this union, the unfilled regions, is the uncertain space.
}
This notion can be extended to hyperrectangles naturally. In a $k$-dimensional hyperrectangle, each Pareto  point can divide the hyperrectangle into $2^k$ sub-hyperrectangles, and those that are Pareto-dominated by \bm{$f^M$} can be discarded.
\cut{Again, if we take union of all these nonviable regions for Pareto optimality, its complement forms the uncertain space.}

Then we have the following result on the effect of finding \bm{$f^M$}.

{\proposition \label{prop:mid_probe} In case that the middle point probe returns no points, we can claim that no Pareto optimal points reside within the sub-hyperrectangle enclosed by its constraints, $\mathcal{C}^{\bm{M}}$, hence reducing the uncertain space. Otherwise, it guarantees to reduce the uncertain space of the current hyperrectangle by discarding the dominated sub-hyperrectangles. } 
\vspace{0.2cm}

\cut{
\begin{proof}
	Denote the volume of a hyper rectangle as $V_{U, N}$ where \bm{$f^U$} and \bm{$f^N$} are the Utopia and Nadir points respectively, \bm{$f^M$} is the middle point calculated by the middle point probe method. Under the continuous assumption, we can see that $f^M_j=(a_j + b_j)/2$ except $j = i$. 
	
	The volume of the original hyper rectangle is: 
	$$V_{U, N}=\prod_{1\leq j \leq k} |f^U_j - f^N_j|.$$	
	
	The volume of the sub hyper rectangle enclosed by the Utopia point and Middle point is:
	
	$$V_{U, M}=\prod_{1\leq j \leq k} |f^U_j - f^M_j| = |f^U_i - f^M_i|*\prod_j |\frac{f^U_i - f^N_i}{2}|$$
	
	where $j \in [i,k]$ and $j \neq i$. Similarly 
	
	$$V_{N, M}=\prod_{1\leq j \leq k} |f^N_j - f^M_j| = |f^N_i - f^M_i|*\prod_j |\frac{f^U_i - f^N_i}{2}|$$
	
	Take the sum we have
	$$V_{U, M} + V_{N, M} = \frac{1}{2^{k-1}}\prod_{1\leq i \leq k} |f^M_i - f^N_i| = \frac{1}{2^{k-1}}*V_{U, N}$$
\end{proof}
}

\minip{A Series of Constrained Optimization Problems.} 
Having explained that a middle point probe defines a CO problem, potentially leading to one Pareto optimal point, 
we next extend our solution  to translate a MOO problem to a series of CO problems in order to cover all (or most) of the Pareto set.

The middle point probe method described above   offers not only a way to find a single Pareto point, but also  a potential iterative method to find more Pareto points. The middle point  \bm{$f^M$} divides the rectangle into $2^k$ sub-hyperrectangles, formed by $k$ hyperplanes that  go through  \bm{$f^M$} and are parallel to the each dimension. 
If we abuse the notation, for each of the sub-hyperrectangles, it is indeed enclosed by its own Utopia point and Nadir point, which can be   calculated from \bm{$f^U}$, \bm{$f^N}$, and \bm{$f^M}$. As seen in Figure \ref{fig:us_example_2d_1}, after finding the middle point \bm{$f^M}$, one can form two sub-hyperrectangles,  enclosed by $(U^1, N^1)$ and $(U^2, N^2)$, respectively. 
Based on this observation, we can derive an iterative method to find more Pareto points by recursively invoking the middle point probe for each sub-hyperrectangle formed.

\minip{Iterative Middle Point Probes.}
Given the initial Utopia and Nadir points, \bm{$f^U$} and \bm{$f^N$},  in $k$-dimensional objective space, the first middle point probe returns one Pareto point, \bm{$f^M$},  that divides the hyperrectangle enclosed by (\bm{$f^U$}, \bm{$f^N$}) into $2^k$ sub-hyperrectangles. The sub-hyperrectangles dominated by \bm{$f^M$} can be safely discarded, while others are added to a queue, each of which has its own local Utopia and Nadir points, ($U^i$, $N^i$), representing the unexplored space with  no knowledge if Pareto points exist.  
For each unexplored sub-hyperrectangle, we remove it from the queue and continue to apply the middle point probe: if the probe returns a new Pareto point, we  generate new sub-hyperrectangles and add them to the queue for exploration later. This process continues until the queue is empty or the maximum number of iterations is reached.

In Figure \ref{fig:us_example_2d_2}, we illustrate the uncertain space given a set of Pareto optimal points in the $2$D case. 
Each of the Pareto points can divide the rectangle enclosed by (\bm{$f^U}$, \bm{$f^N}$) into four parts, and the union of the colored regions is where no Pareto optimal points exist. The complement of this union, the unfilled regions, is the uncertain space.
In a $k$-dimensional hyperrectangle, again, if we take union of all those nonviable regions for Pareto optimality, its complement forms the uncertain space.

Following the previous propositions, we have the following result on our returned solution set.

{\proposition \label{prop:conclusion} If we start the Iterative Middle Point Probes procedure from the initial Utopia and Nadir points, and let it terminate until the uncertainty space becomes empty, then in the 2D case, our procedure guarantees to find all the Pareto points if they are finite. In high-dimensional cases, it  guarantees to find a subset of  Pareto optimal points.}
\vspace{0.2cm}

\minip{Filtering of Candidate Points}. As explained previously, in general it is not guaranteed that the CO solutions returned are Pareto optimal in high-dimensional cases. Therefore, we can add a filter at the end of our Progressive Frontier approach to remove the solutions that are not Pareto optimal. This is similar to the NC method~\cite{messac2003nc}.


%% file: algorithms.tex
\section{PF Algorithms}
In this section, we present algorithms that implement the Progressive Frontier (PF) approach. 
We note that most MOO algorithms suffer from exponential complexity with respect to the number of the objectives, $k$. This is because the number of non-dominated points tends to grow very quickly with $k$ and the time complexity of computing the volume of dominated space grows super-polynomially with $k$~\cite{Emmerich:2018:TMO}. For this reason, the MOO literature refers to  optimization with up to 3 objectives as {\bf multi-objective} optimization, whereas  optimization with more than 3 objectives is called {\bf many-objective} optimization and handled using different methods such as
preference modeling~\cite{Emmerich:2018:TMO} or fairness among different objectives~\cite{tan2016tempo}. 

Since our goal is to develop a practical solution for a cloud optimizer, most of our use cases fall in the scope of multi-objective optimization. However, a unique requirement in our problem setting is to compute the Pareto frontier in a few seconds,  which hasn't been considered previously~\cite{Emmerich:2018:TMO}. To achieve our performance goal, we present several techniques, including {\em uncertainty-aware incremental computation}, {\em fast approximation}, and {\em parallel computation}.

\subsection{A Deterministic Sequential Algorithm}
\label{subsec:sequential}
We first present a deterministic, sequential algorithm that implements  the Progressive Frontier (PF) approach, referred to as \pfs. This algorithm has two key features:

First, it is {\em incremental}, in the sense that  it first constructs a  Pareto frontier $\tilde{\mathcal{F}}$  by using a small number of points and expands $\tilde{\mathcal{F}}$ with more points as more time is invested. 
This feature is crucial because finding one Pareto point  is already expensive due to  being a mixed-integer nonlinear programming problem~\cite{bussieck2003minlp,Liberti2018}. Hence, one cannot expect the optimizer to find all Pareto points at once. Instead, it  produces $n_1$ points first (e.g., those that can be computed within the first second), and then expands the frontier with additional $n_2$ points, afterwards $n_3$ points, and so on. 

Second, this algorithm is {\em uncertainty-aware}, returning more points in the regions of the Pareto frontier that lack sufficient information. \cut{In practice, we need to  consider  time constraints in running the PF approach to enable timely  recommendation of configurations. }If the Pareto frontier includes a large number of points,  under a time constraint we can return only a subset of them as approximation.
In this case, we would like to generate the next set of CO problems such that the points returned bring more valuable information.
\cut{In contrast, the NC method attempts to cover the objective space evenly, but it does not leverage the information already available in the frontier; for example, some region of the frontier is flat and will not gain new information with additional points, whereas another region includes a sharp turn and hence needs more points to better characterize the turning point. }

To do so, we want to augment the Iterative Middle Point Probe method (\S\ref{subsec:progressive}) 
by further controlling how to choose the {\bf best} sub-hyperrectangle to explore next, among those that have not been explored. We do so by defining a measure, the {\em volume of uncertain space}, to capture how much the current frontier $\tilde{\mathcal{F}}$ may deviate from the true yet unknown frontier $\mathcal{F}$. (Note that both $\tilde{\mathcal{F}}$ and  $\mathcal{F}$ may include an infinite set of points.)
The volume of uncertain space can be calculated from a related set of sub-hyperrectanges, which allows us to rank the sub-hyperrectangles that have not been explored.
Among those, the sub-hyperrectangle with the largest volume of uncertain space  will be chosen to explore next.

Now we are ready to present the complete sequential  algorithm. It is an incremental method trying to reduce the uncertain space as fast as possible.  Below we give a brief description of it, while the details are shown in Algorithm~\ref{alg:greedy}.

\textbf{Init:} Find the reference points by solving $k$ single objective optimization problems. Form the initial Utopia and Nadir ($U^0$ and $N^0$) points, and construct the first hyperrectangle. Prepare a priority queue in decreasing order of hyperrectangle volume, initialized with the first hyperrectangle. 

\textbf{Iterate:} Pop a hyperrectangle from the priority queue. Apply the middle point probe to find a Pareto point, \bm{$f^M$}, in the current hyperrectangle, which is formed by $U^i$ and $N^i$ and with {\bf largest} volume among all the existing hyperrectangles. Divide the current hyperrectangle into $2^k$ sub-hyperrectangles, discard those that are dominated by \bm{$f^M$}, and calculate the volume of the others. Put them in to the priority queue. 

\textbf{Terminate:} when the desired number of solutions is reached.

\textbf{Filter:} Check the result set, and remove any point dominated by another one in the result set.

\begin{algorithm}[t]
	\caption{Progressive Frontier-Sequential (PF-S)}
	\label{alg:greedy}
	\small
	\begin{algorithmic}[1]
		
		\REQUIRE {$k$ lower bounds(LOWER): $lower^j$, \\$k$ upper bounds(UPPER): $upper^j$, \\number of points: $M$}
		\STATE $PQ \gets \phi$    \COMMENT{PQ is a priority queue sorted by hyperrectangle volume}
		\STATE $plan_i \gets optimize^i(LOWER, UPPER)$    \COMMENT{Single Objective Optimizer takes LOWER and UPPER as constraints and optimizes on $i$th objective}
		\STATE $Utopia, Nadir \gets computeBounds(plan_1, \ldots, plan_k)$
		\STATE $volume \gets$ computeVolume$(Utopia, Nadir)$
		\STATE $seg \gets (Utopia, Nadir, volume)$
		\STATE $PQ$.put($seg$)
		\STATE $count \gets k$
		
		\REPEAT
		\STATE $seg \gets$ $PQ$.pop()
		\STATE $Utopia \gets seg.Utopia$
		\STATE $Nadir \gets seg.Nadir$
		\STATE $Middle \gets (Utopia + Nadir)/2$
		\STATE $Middle_i \gets optimize^i(Utopia, Middle)$ \COMMENT{Constraint Optimization on $i$-th objective}
		\STATE $\{plan\} \gets Middle_i$
		\STATE $count+=1$
		
		\STATE $\{rectangle\} = generateSubRectangles(Utopia, Middle, Nadir)$ \COMMENT{return $2^k - 1$ rectangles, represented by each own Utopia and Nadir}
		
		\FOR{each rectangle in \{rectangle\}}
		\STATE $Utopia \gets rectangle.Utopia$    
		\STATE $Nadir \gets rectangle.Nadir$ 
		\STATE $volume \gets$ computeVolume$(Utopia, Nadir)$
		\STATE $seg \gets (Utopia, Nadir, volume)$
		\STATE $PQ$.put($seg$)
		\ENDFOR  
		\UNTIL{$count > M$}
		
		\STATE $output \gets filter(\{plan\})$\COMMENT{remove plan dominated by another plan in the same set}
	\end{algorithmic}
\end{algorithm}

\begin{figure*}[t]
	\centering
	\hspace{-12cm}
	
	\begin{tabular}{llll}
		
		\multicolumn{3}{l}{
		\subfigure[\small{Constrained optimization with multiple objectives (e.g., $F_1$ is modeled by a DNN and $F_2$ by a GP)}]
		{\label{fig:co-solver}\includegraphics[height=3.2cm,width=12cm]{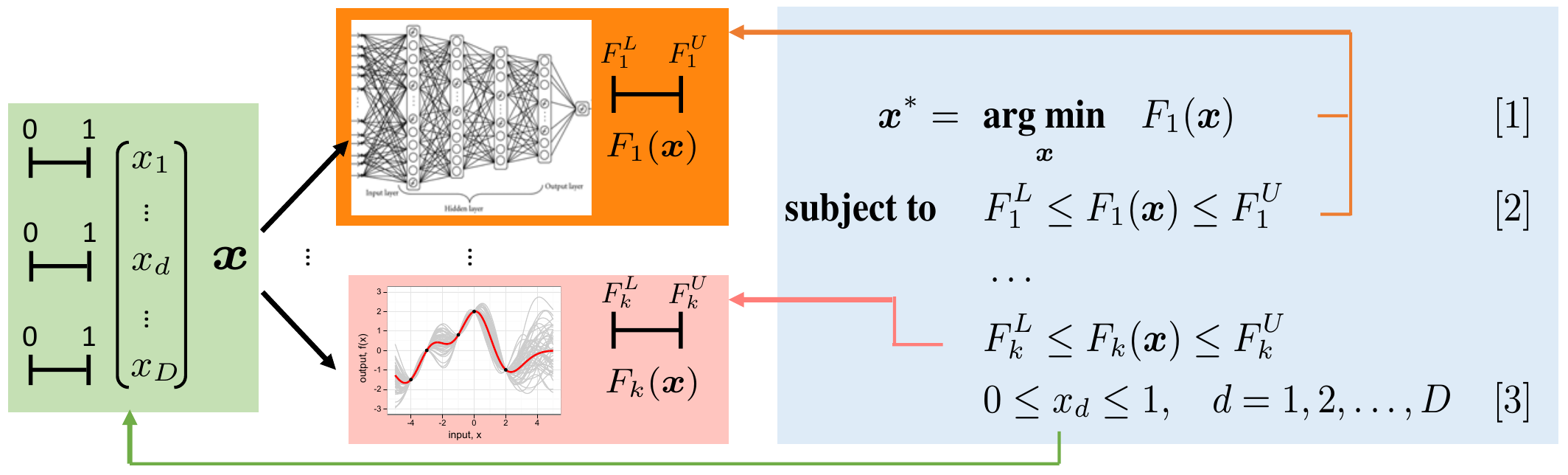}}
		}
		
		&
		\subfigure[\small{Loss term on obj1, $\phi_1=\hat{F_1}$}]
		{\label{fig:loss-obj1}\includegraphics[height=3.0cm,width=3.8cm]{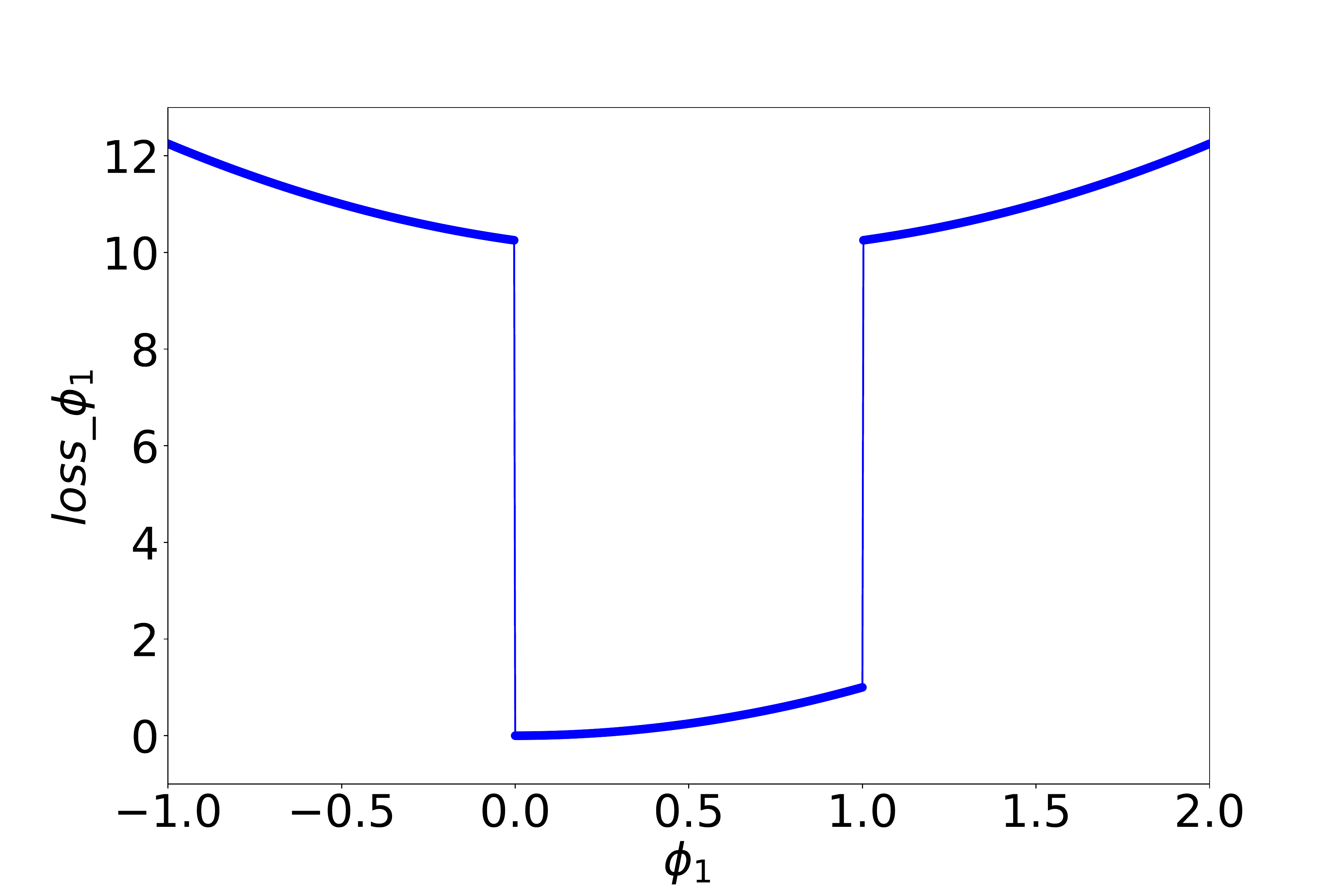}}
		
		\\	
		
		\subfigure[\small{Loss term on obj2, $\phi_2=\hat{F_2}$}]
		{\label{fig:loss-obj2}\includegraphics[height=3.0cm,width=3.7cm]{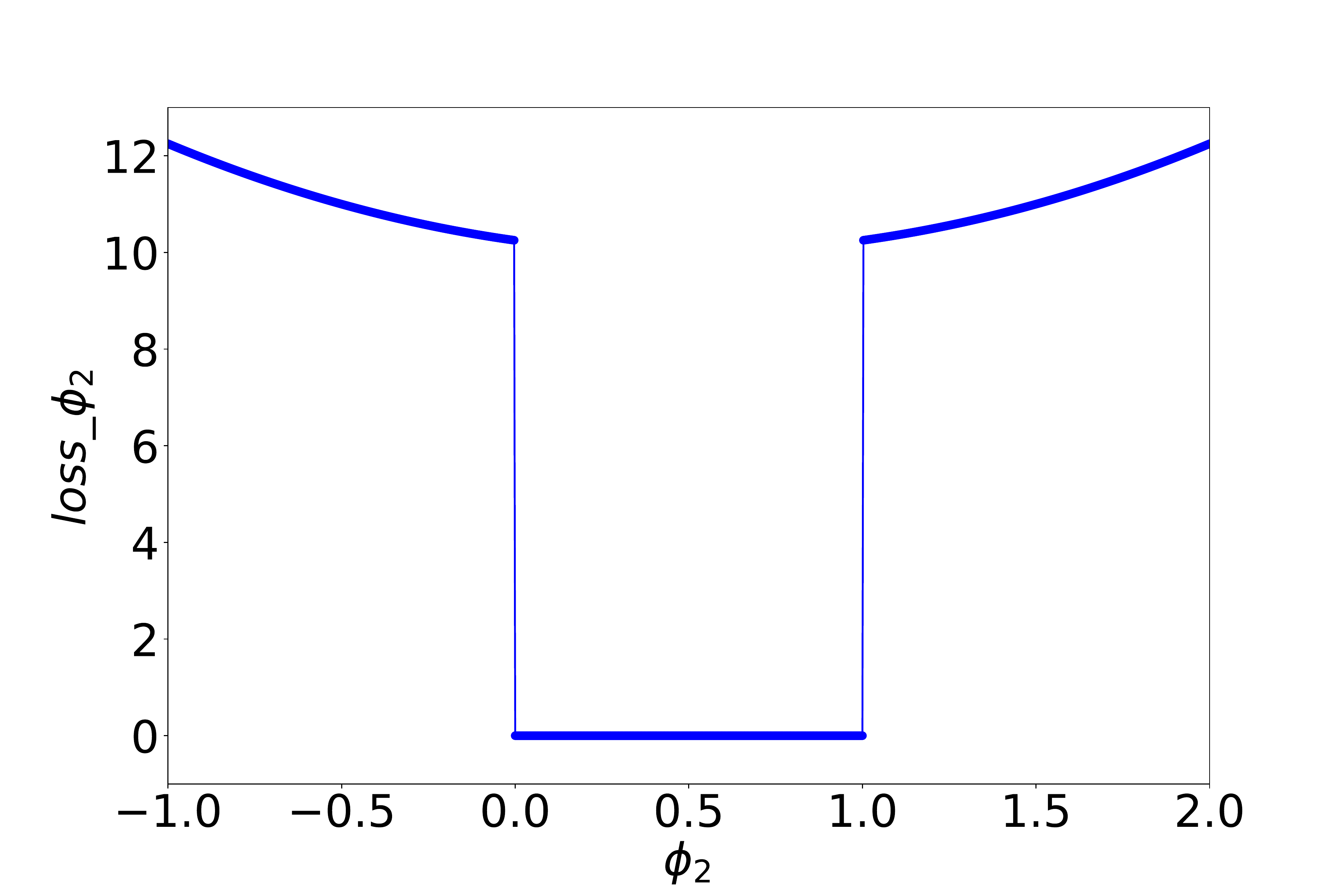}}

		&
		\subfigure[\small{$\mathcal{L}(x)$ on univariate input $x$, $F_1=max(0,12x-3)$, and $F_2=max(0,8x-1)$}]
		{\label{fig:loss-x}\includegraphics[height=3.0cm,width=3.7cm]{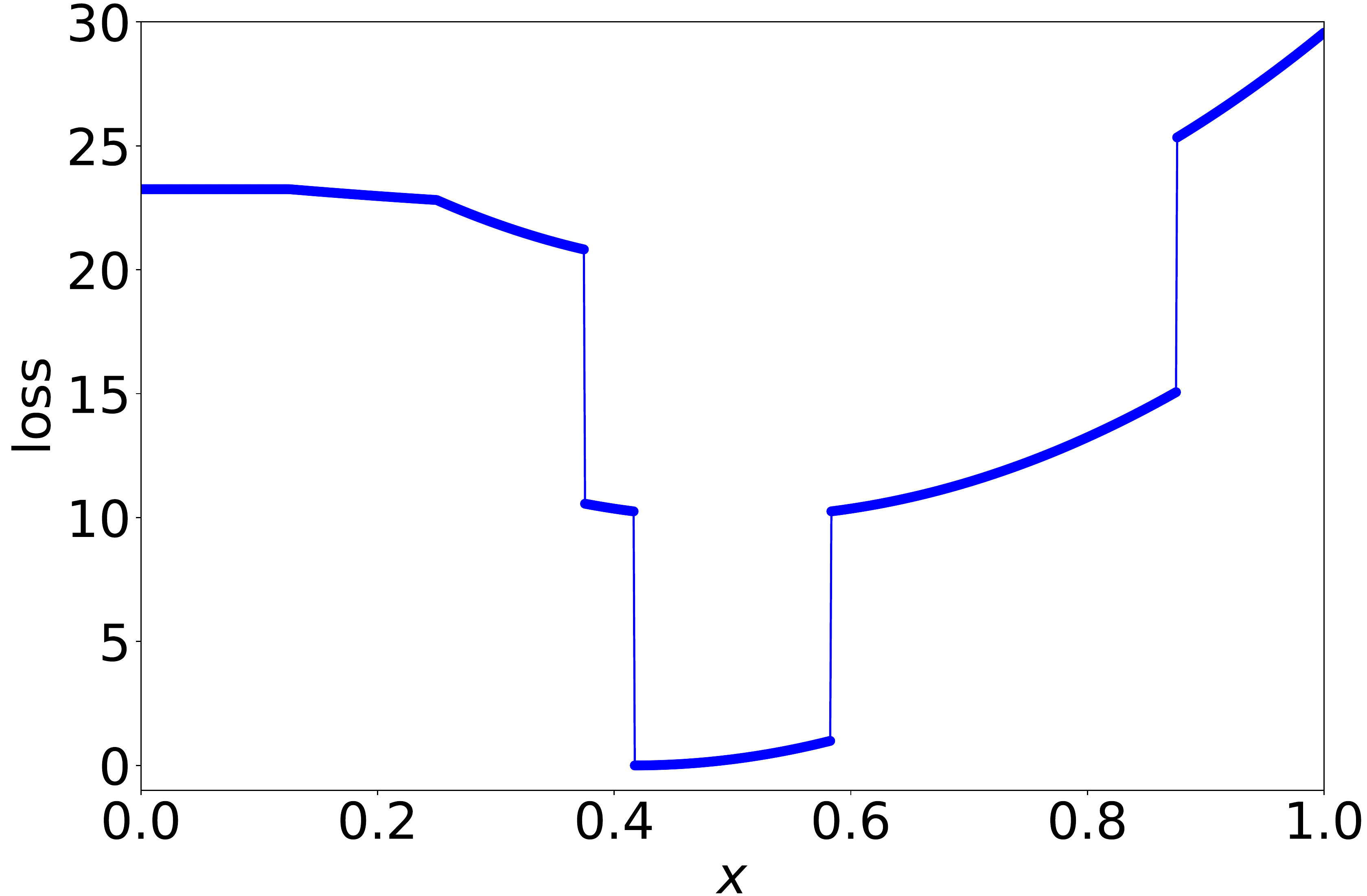}}

		&
		\subfigure[\small{$\mathcal{L}(x_1,x_2)$ on bivariate input $x_1$ and $x_2$, $F_1=max(0,12x_1+8x_2-8)$, and $F_2=max(0,4x_1-16x_2+4)$}]
		{\label{fig:loss-x1x2}\includegraphics[height=3.2cm,width=5.2cm]{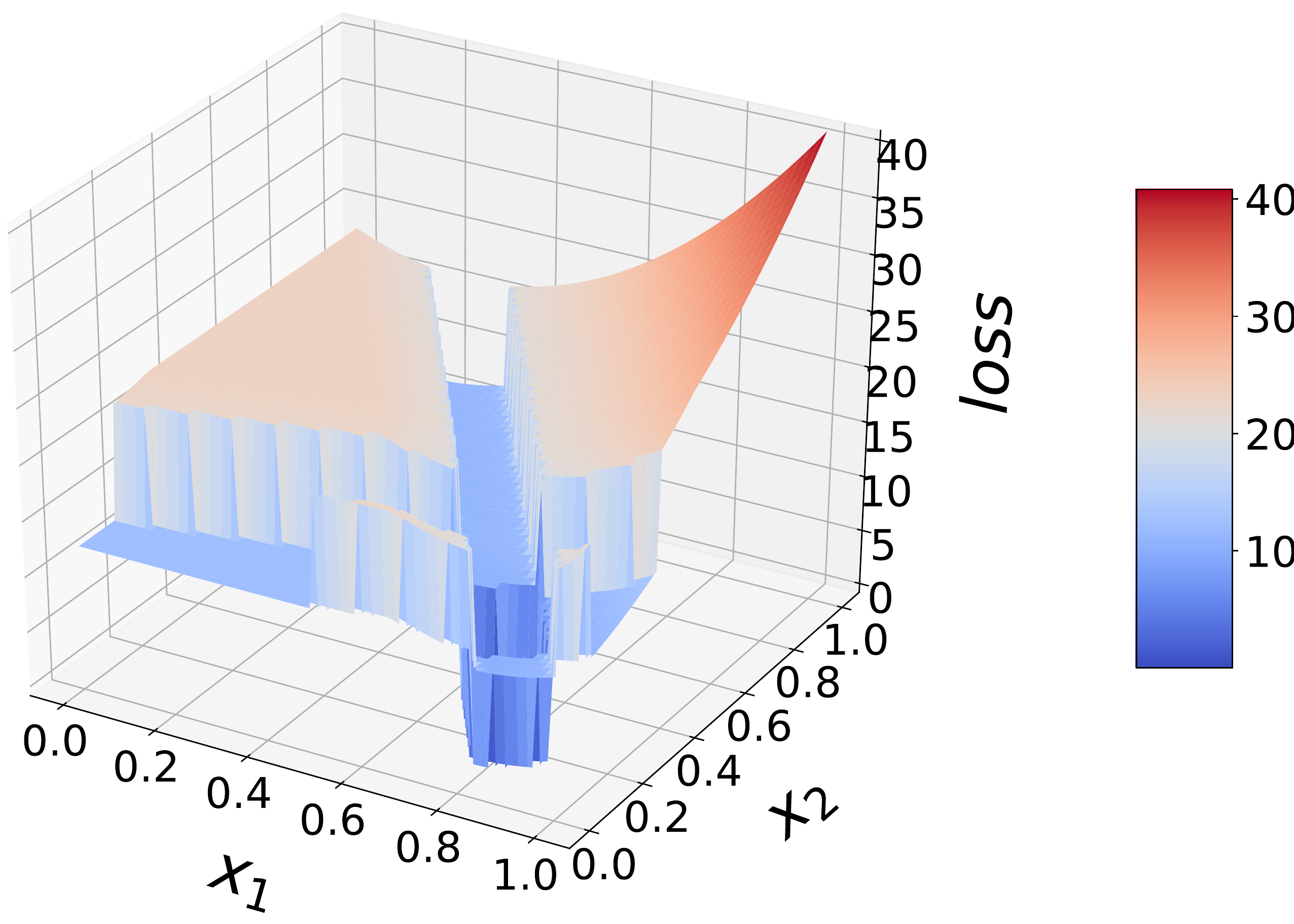}}

		&
		\subfigure[\small{A GP fn on $x$ with expected values (the blue line) and model uncertainty (the pink region)}]
		{\label{fig:gpr}\includegraphics[height=3.0cm,width=3.8cm]{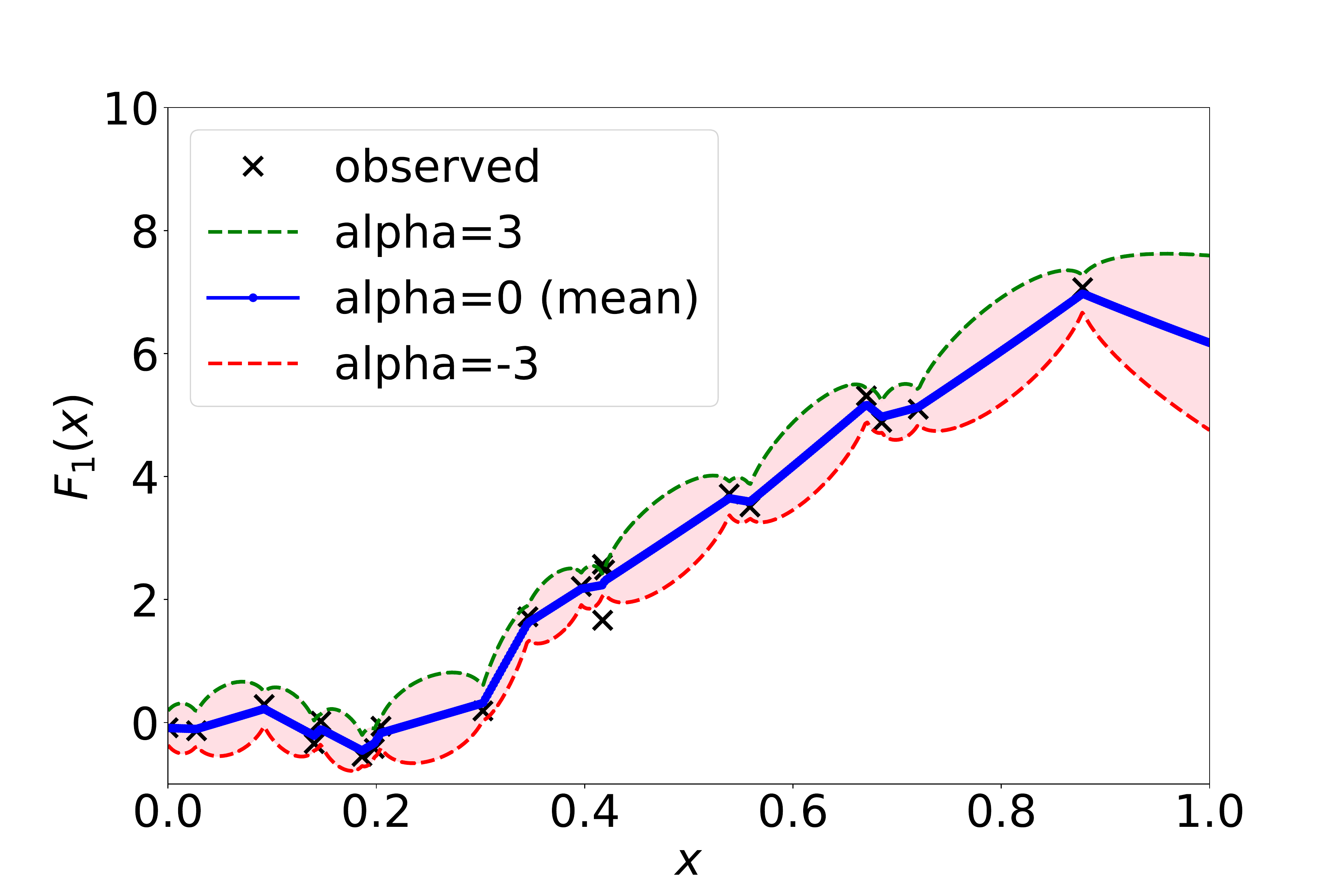}}

	\end{tabular}
\vspace{-0.2in}
\caption{\small Constrained optimization with multiple objectives, and the loss function, $\mathcal{L}$, on different objectives and input parameters $\bm{x}$}
 \label{fig:moo}
\vspace{-0.1in}
\end{figure*}

%
%
%
%

\subsection{Multi-Objective Gradient Descent}
\label{subsec:dnn}

We next consider an important subroutine, $optimize()$, that solves each constrained optimization problem (line 13 of Algorithm~\ref{alg:greedy}). 
Recall that our  objective functions are given by learned models, $\Psi_i(\bm{x})$, $i=1\ldots k$, where each model is likely to be non-linear and some variables among $\bm{x}$ can be integers.  Even when restricted to a single objective, this problem reduces to a mixed-integer nonlinear programming (MINLP) and is  NP-hard~\cite{bussieck2003minlp,Liberti2018}. There is not one general MINLP solver that will work effectively for every  nonlinear programming problem~\cite{NEOS-guide}.
For example, many of the MINLP solvers~\cite{NEOS-server} fail to run because they assume certain properties of the objective function $F$, e.g., twice continuously differentiable (Bonmin~\cite{Bonmin}) or  factorable into the sumproduct of univariate functions (Couenne~\cite{Couenne}), 
which do not suit our learned models represented such as Deep Neural Networks (DNNs).
The most general MINLP solver, Knitro~\cite{Knitro}, runs for our learned models but very slowly, e.g., 42 minutes for solving a single-objective optimization problem when the learned model is a DNN,  
or 17 minutes when the model is a Gaussian Process (GP). 
Such a solution is too slow for us to use even for single-objective optimization, let alone the extension to multiple objectives.

In this work, we propose a novel solver that employs a customized gradient descent approach to approximately solve our constrained optimization (CO) problems  involving multiple objective functions. 
This CO problem is illustrated in Figure~\ref{fig:co-solver}.


In the first step, we transform variables to prepare for optimization by following the common practice in machine learning:  
Let $\mathbf{x}$ be the original set of parameters, which can be categorical, integer, or continuous variables. 
If a variable is categorical , we use one-hot encoding to create dummy variables. For example, 
if $x_d$ takes values \{a, b, c\}, we create three boolean variables, $x_d^a$, $x_d^b$, and $x_d^c$, among which only one takes the value `1'. Afterwards all the variables are normalized to the range [0, 1], and boolean variables and (normalized) integer variables are relaxed to continuous variables in [0, 1]. As such, the CO problem deals only with continuous variables in [0,1], which we denote as $\bm{x}=x_1, \ldots, x_{D} \in [0,1]$. After a solution is returned for the CO problem, we set the value for a categorical attribute based on the dummy variable with the highest value, and round the solution returned for a normalized integer variable to its closest integer. Our work also employs an optimization to  support categorical variables using  parallel processing, which we defer to the next subsection.

Next, we focus on the CO problem. 
Our design of a {\bf Multi-Objective Gradient Descent   (MOGD) solver} uses carefully-crafted loss functions to guide gradient descent to find the minimum of a target objective while satisfying a variety of constraints, where both the target objective and constraints can be specified over complex models, e.g., using DNNs, GPs, or other regression functions.

1. {\em Single objective optimization}. 
As a base case, we consider single-objective optimization, \textit{minimize} $F_1(\bm{x}) = \Psi_1(\bm{x})$. 
For optimization, we set the loss function simply as, $\mathcal{L}(\bm{x}) = F_1(\bm{x})$.
Then starting from an initial configuration, $\bm{x}^0$, gradient descent  (GD) will iteratively adjust the configuration to a sequence $\bm{x}^1, \ldots, \bm{x}^n$ in order to minimize the loss, until it converges to a local minimum or reaches  a maximum of steps. To increase the chance of hitting a global minimum, we use a {\em multi-start} method to try gradient descent from multiple initial points, and finally choose $\bm{x}^\ast$ that gives the smallest value among these trials. Finally, to cope with the constraint on each variable, 
$0 \leq x_d \leq 1$, we restrict the GD process such that 
when it tries to walk across the boundary of $x_d$, we simply set $x_d$ to the boundary value. In future iterations, GD may not be able to reduce the loss by pushing $x_d$ outside the boundaries, but it can still adjust other variables until reaching the stopping criteria. 

2. {\em Constrained optimization}. Next we consider a constrained optimization (CO) problem, as shown in  Figure~\ref{fig:co-solver}. Without loss of generality, we treat $F_1$ as the target objective and  $F_j \in [F_j^L, F_j^U]$, $j=1,\ldots, k$ as constraints.  Recall that these constraints [$F_j^L, F_j^U$] come from our PF algorithm that at each step, aims to explore a specific region (a hyper-rectangle formed by these constraints) in the objective space by solving a particular CO problem. 

To solve the CO problem, we need to design an appropriate loss function,  $\mathcal{L}(\bm{x})$, such that by minimizing this loss, we can minimize $F_1(\bm{x})$ while at the same time satisfying all the constraints. 
Our proposed loss function is as follows: 
\begin{equation}
    \begin{split}
    \mathcal{L}(\bm{x})  = \hspace{0.2cm} & \mathbbm{1} \{0 \leq \hat{F_{1}}(\bm{x}) \leq 1 \} \cdot \hat{F_{1}}(\bm{x})^2 + \\
     \sum_{j=1}^k & \mathbbm{1}\{\hat{F_j}(\bm{x})>1 \lor \hat{F_j}(\bm{x})<0\} [(\hat{F_j} (\bm{x}) - \frac{1}{2})^2 + P] \\
    \end{split}
    \label{co-loss}
\end{equation}
where $\hat{F}_j(\bm{x}) = \frac{F_j(\bm{x}) - F_j^L}{F_j^U - F_j^L}$, for $j \in [1, k]$, denotes the normalized value of each objective,  and $P$ is a constant for extra penalty.  Since the range of each objective function $F_j(\bm{x})$ varies, we first normalize each objective according to its upper and lower bounds, so that $\hat{F}_j^L=0$, $\hat{F}_j^U=1$, and a valid objective value $\hat{F}_j(\bm{x}) \in [0, 1]$. 

Figures~\ref{fig:loss-obj1}-\ref{fig:loss-obj2} illustrate the breakdown of the terms of the above loss when we have a single objective $F_1$ and constraints on objectives $F_1$ and $F_2$.
The loss for $F_1$ has two terms. When $F_1$ falls in the region ($0 \leq \hat{F}_1(\bm{x}) \leq 1$), the first term of the loss penalizes a large value of $F_1$, and hence minimizing the loss helps reduce $F_1$. The second term of the loss aims to push the objective into its constraint region. If $F_1(\bm{x})$ cannot satisfy the constraints ($\hat{F}_j(\bm{x}) > 0 \lor \hat{F}_j(\bm{x}) < 1$), it will contribute a loss according to its distance from the valid region. The extra penalty, $P$, is a large constant added to ensure that the loss for $F_1$ if it falls outside its valid region is much higher than that if $F_1$ lies in its valid region. Figures~\ref{fig:loss-obj1} shows the combined effect of these two terms over $F_1$. 
In comparison, the loss for $F_2$ has only the second term, i.e., to push it to satisfy its constraint, which is shown in Figure-\ref{fig:loss-obj2}.
The final loss combines the terms for both $F_1$ and $F_2$. 

Figure~\ref{fig:loss-x} illustrates the loss function, $\mathcal{L}$, over univariate input $x$, assuming two specific models for $F_1$ and $F_2$, each of which is a simple neural network with ReLU as the activation function, where 
Figure~\ref{fig:loss-x} shows the loss over bivariate input $x_1$ and $x_2$. 
Such loss will be used to guide gradient descent such that by minimizing $\mathcal{L}$, it is likely to find the values of the input variables that minimize the target objective while satisfying the constraints. 
A final note is that GD usually assumes  the loss function to be differentiable, but our loss function is not at specific points. However, we can use {\em subderivative}: for a point $\bm{x}^0$ that is not differentiable, we can choose a value between its left derivative and right  derivative, which is  called a subgradient. Machine learning libraries allow subgradients to be defined by the user program and can automatically handle common cases, including our loss functions for DNNs and GPs.

3. {\em Handling model uncertainty}. 
We have several extensions of our MO-GD solver. In the interest of the space, we highlight the most important extension, that is, to support {\em model uncertainty}. Since our objective functions use learned models, these models may not be accurate before sufficient training data is available. Hence, it is desirable that the optimization procedure take into account model uncertainty when recommending an optimal solution. Fortunately, advanced machine learning techniques can support a regression task, $F(\bm{x})$, with both expected value $\mathbb{E}[F(\bm{x})]$ and variance, $\mathtt{Var}[F(\bm{x})]$. Such techniques include Gaussian Processes~\cite{SchulzSK2018}, with an example shown in Figure~\ref{fig:gpr}, and Bayesian approximation for DNNs~\cite{gal16}. 
Given such information, we only need to replace each objective function, $F_j(\bm{x})$, with 
$\tilde{F}_j(\bm{x}) = \mathbb{E}[F_j(\bm{x})] + \alpha \cdot \mathtt{std}[F_j(\bm{x})]$, where $\alpha$ is a small positive constant. 
As such, $\tilde{F}_j(\bm{x})$ offers a more conservative estimate of $F_j(\bm{x})$ for solving a minimization problem, given the model uncertainty. Then we use $\tilde{F}_i(\bm{x})$ to build the loss function as  in Eq.~\ref{co-loss} to solve the CO problem.

Finally, we note three sources of approximation in our MO-GD solver: (1) the relaxation of categorical and integer variables to continuous variables in [0,1]; (2) the use of  Gradient Descent (GD)  to solve a non-convex optimization problem, where solutions are likely to be local minima; (3) the model uncertainty. As our benchmark results show, models learned by OtterTune~\cite{VanAken:2017:ADM} can have prediction errors $\in [10\%, 40\%]$ compared to observed values. Thus, model uncertainty has a major impact on the  performance achieved by the solution  from the MO-GD solver. This indicates that to build a practical system, it is crucial  that the model be updated frequently from new training data and our multi-objective optimizer refresh the Pareto frontier based on the new model in order to make more accurate recommendations -- for this reason, we treat the speed of computing the Pareto frontier as a key performance goal. 

\subsection{Approximate and Parallel PF Algorithms}  

{\bf Approximate Sequential Algorithm} (\sdnn): 
When we implement single-objective optimization (Line 2 of Algorithm~\ref{alg:greedy}) and constrained optimization (Line 13) using our MO-GD solver, we obtain a new algorithm called PF Approximate Sequential. 
Note that this leads to an approximate Pareto set for the MOO problem because each solution of a CO problem can be suboptimal. In fact, the most powerful commercial solver, Knitro~\cite{Knitro}, also returns approximate solutions due to the complex, non-convex properties of our objective functions, despite its long running time. 

{\bf Approximate Parallel Algorithm} (\pdnn): 
We finally present a parallel version of the  approximate algorithm,  PF-Approximate Parallel (\pdnn). 
The main difference from the approximate sequential algorithm is  that given any hyperrecetange we aim to explore at the moment, we partition it into a $l^k$ grid and for each grid cell, construct a CO problem using the 
the Middle Point Probe (Eq.~\ref{eq:mpp}). We send these $l^k$ CO  problem to our MO-GD solver simultaneously. Internally, our solver will solve these problems in parallel (using a multi-threaded implementation). Some of the cells will not return any Pareto point and hence will be omitted. For each of those cells that returns a Pareto point, the Pareto point breaks the cell into a set of sub-hyperrectangles that need to be further explored, and are added to a priority queue as before. Afterwards, the  sub-hyperrectangle with the largest volume is removed from the queue. To explore it, we further partition it into $l^k$ cells and ask the solver to solve their corresponding CO problems simultaneously. This process terminates when the queue becomes empty.

{\bf Parallel Models.} Our work also supports categorical attributes using parallel processing. As OtterTune~\cite{VanAken:2017:ADM} suggests, we can perform feature selection to focus on a small set ($\sim$10) of variables that impact a specific model the most. If among the selected variables, there are only a limited number of categorical values to consider, e.g., a boolean variable indicating compressing immediate data or not, we can build value-specific models, one for each categorical value, and compute them using multiple threads simultaneously.

%% file: recommendation.tex


\section{Automatic Solution Selection}
\label{subsec:recommendation}

Once a Pareto set is computed for a workload, our optimizer employs a range of strategies to recommend a new configuration from the set. We highlight the most effective strategies below and describe other recommendation strategies  in \techreport{Appendix~\ref{appendix:selection}}{our technical report~\cite{Song2020}}.

First, the \textbf{Utopia Nearest} (UN) strategy chooses the Pareto point closest to the Utopia point \bm{$f^U$}, by computing the Euclidean distance of each point in the Pareto set $\tilde{\mathcal{F}}$ to \bm{$f^U$} and returning the point that minimizes the distance. 

A variant  is the \textbf{Weighted Utopia  Nearest} (WUN) strategy, which uses a weight vector, $\bm{w}=(w_1,...w_k)$, to capture the importance among different objectives and is usually set based on the application preference. A further improvement is workload-aware WUN, motivated by our observation that  expert knowledge about different objectives is available from the literature. For example, between latency and cost, it is known from the literature of Parallel Databases~\cite{DeWittG92} that it is beneficial to allocate more resources to large queries (e.g. join queries) but less so for small queries (e.g., selection queries). In this case, we use  historical data to divide workloads into three categories, (low, medium, high), based on the observed latency under the default configuration. For long running workloads, we give more weight to latency than the cost, hence encouraging more cores to be allocated; for short running workloads, we give more weight to the cost, limiting the cores to be used. We encode such expert knowledge using internal weights, $\bm{w}^I=(w_1^I, ..., w_k^I)$, and application preference using external weights, $\bm{w}^E$$=(w_1^E, ..., w_k^E)$. The final weights are $\bm{w}=(w_1^I w_1^E, ..., w_k^I w_k^E)$.


\cut{
\begin{algorithm}[H]
 \caption{Maximum Slop Algorithm: $maxSlop$}
\label{alg:max_slop}
\small
\begin{algorithmic}[1]
\REQUIRE {Pareto optimal points set: $PF$, \\Reference point: $r$}
\STATE $slop_max \gets 0$
\STATE $optimal \gets $ NULL
\FOR{each $p \in PF$}
    \STATE $s \gets$ slop($p$, $r$)
    \IF {$s > slop_{max}$ }
        \STATE $slop_{max} \gets s$
        \STATE $optimal \gets p$
    \ENDIF
\ENDFOR
\RETURN $optimal$
\end{algorithmic}
\end{algorithm}
}

\cut{
\begin{algorithm}[H]
 \caption{Knee Point Algorithm: $kneePoint$}
\label{alg:knee_point}
\small
\begin{algorithmic}[1]

\REQUIRE {Pareto optimal points set: $PF$}
\STATE $ratio_max \gets 0$
\STATE $optimal \gets $ NULL
\STATE $ul \gets PF$.first()
\STATE $lr \gets PF$.last()
\FOR {each $p \in PF \backslash ul\backslash lr$}
    \STATE $s_l \gets$ slop($ul$, $p$)
    \STATE $s_r \gets$ slop($p$, $lr$)
    \STATE $ratio \gets s_l/s_r$
    \IF {$ratio > ratio_{max}$}
        \STATE $ratio_{max} \gets ratio$
        \STATE $optimal \gets p$
    \ENDIF
\ENDFOR
\RETURN $optimal$
\end{algorithmic}
\end{algorithm}
}

%% file: experiments.tex
\section{Performance Evaluation}
\label{sec:experiments}

\begin{figure*}[t]
	\centering
	\vspace{-0.3in}
	\hspace{-6cm}
	
	\begin{tabular}{lcc}

		\subfigure[\small{Uncertain space (job 9, 2d)}]
		{\label{fig:batch-9-3-time}\includegraphics[height=3.2cm,width=5.5cm]{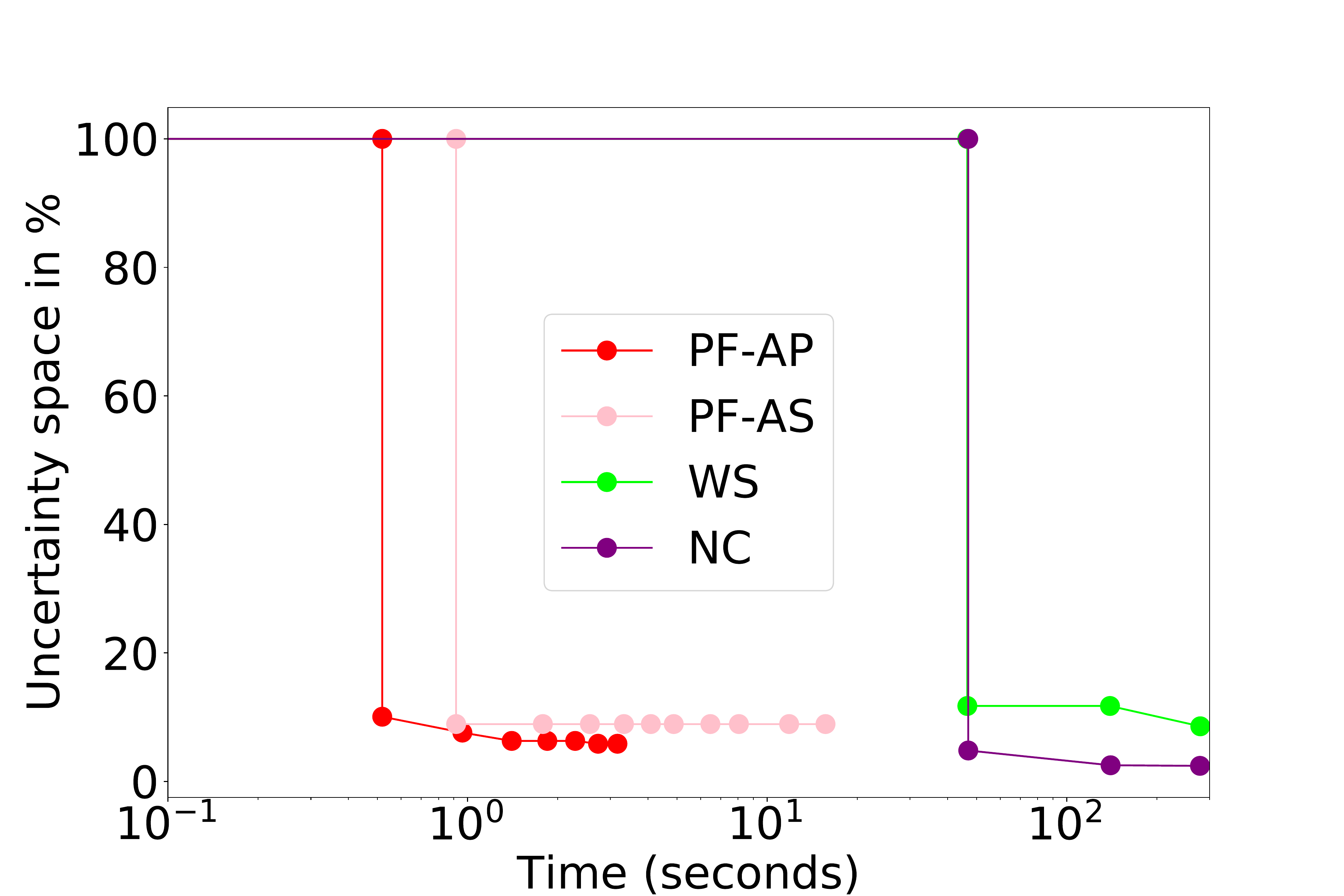}}

		&	
		\subfigure[\small{Frontier of WS and NC (job 9, 2d)}]
		{\label{fig:batch-9-3-ws-nc}\includegraphics[height=3.2cm,width=5.5cm]{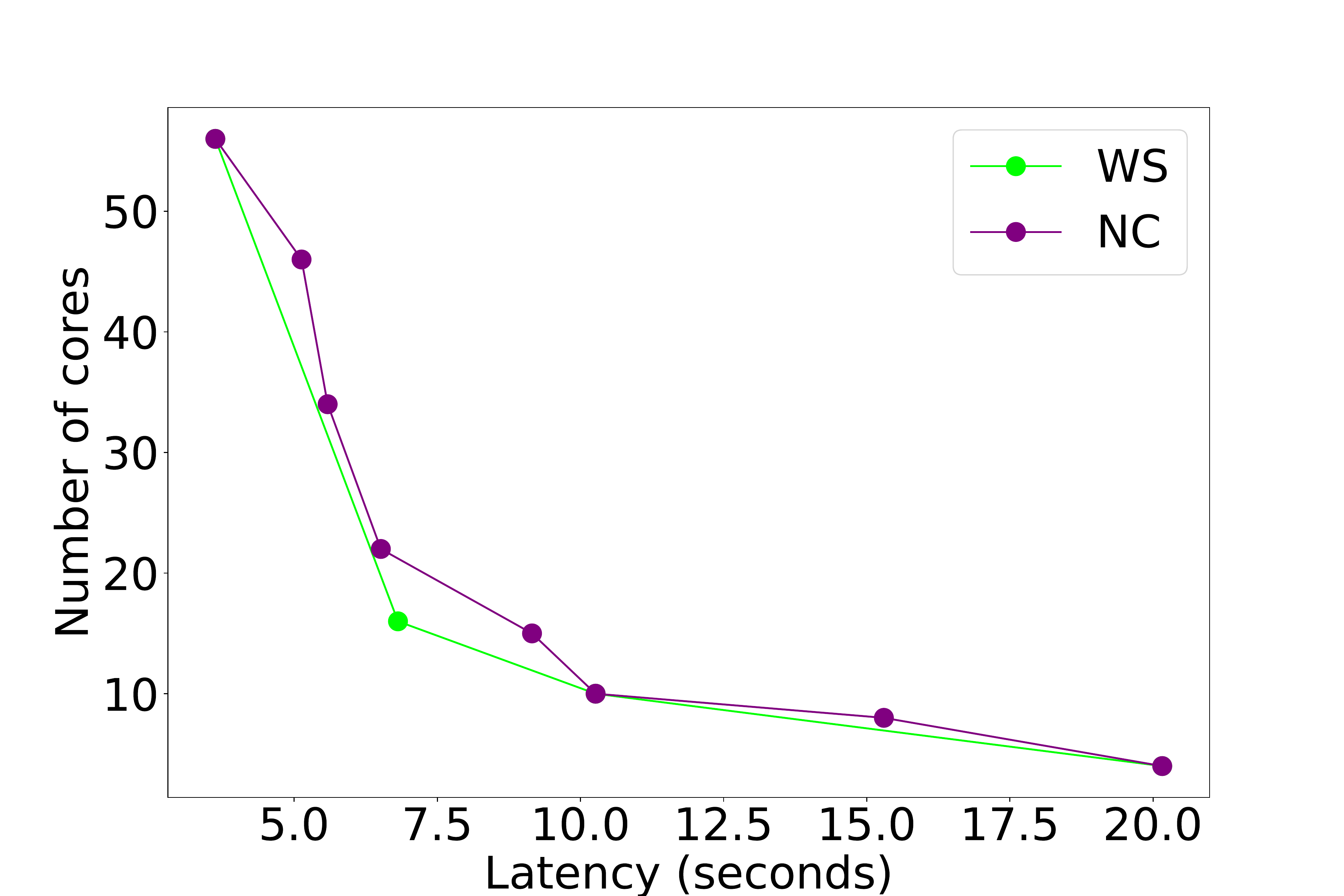}}
	
		&	
		\subfigure[\small{Frontier of PF (job 9, 2d)}]
		{\label{fig:batch-9-3-pf}\includegraphics[height=3.2cm,width=5.5cm]{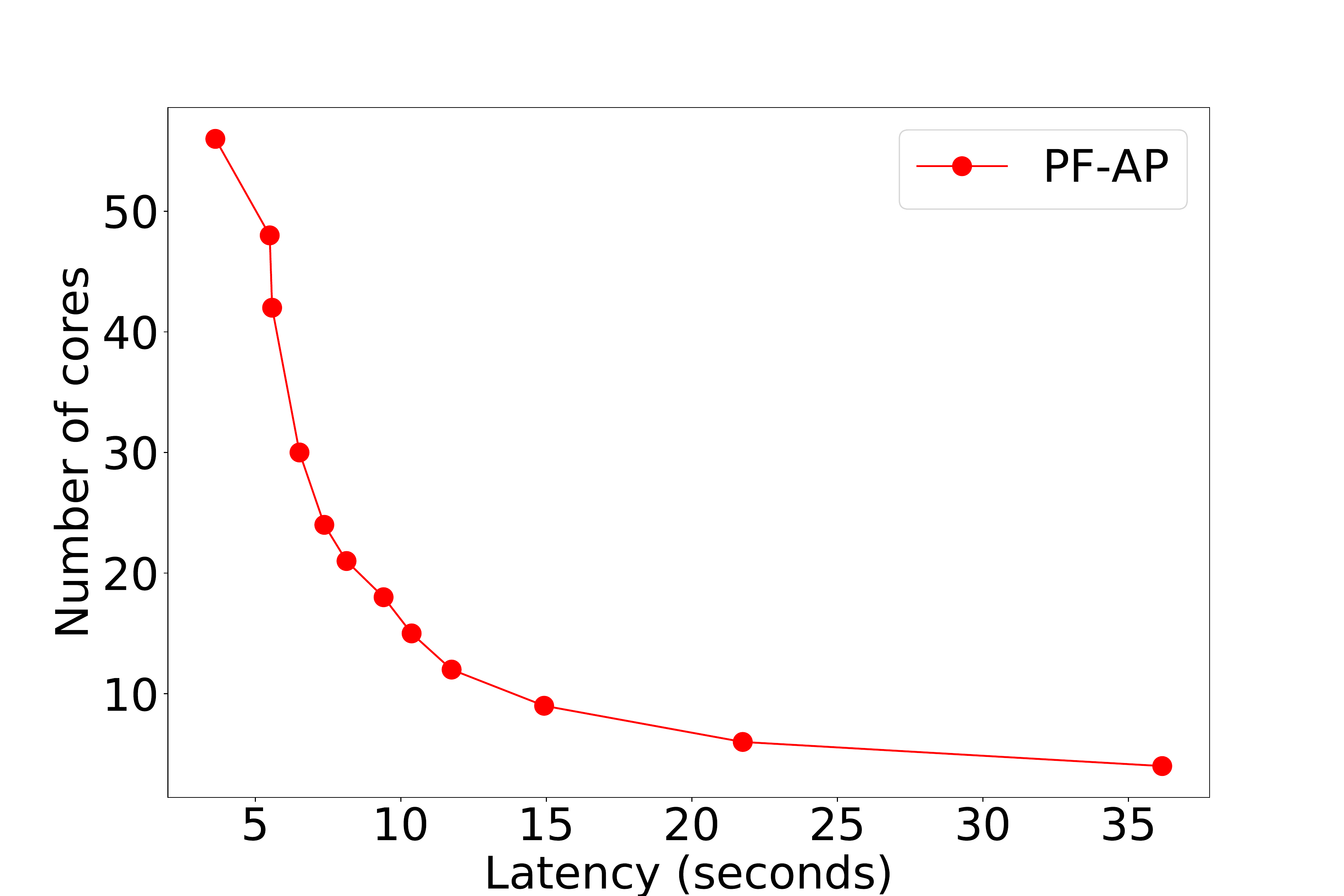}}

		\\	
			
		\subfigure[\small{Uncertain space (job 9, 2d)}]
		{\label{fig:batch-9-3-time-evo}\includegraphics[height=3.2cm,width=5.5cm]{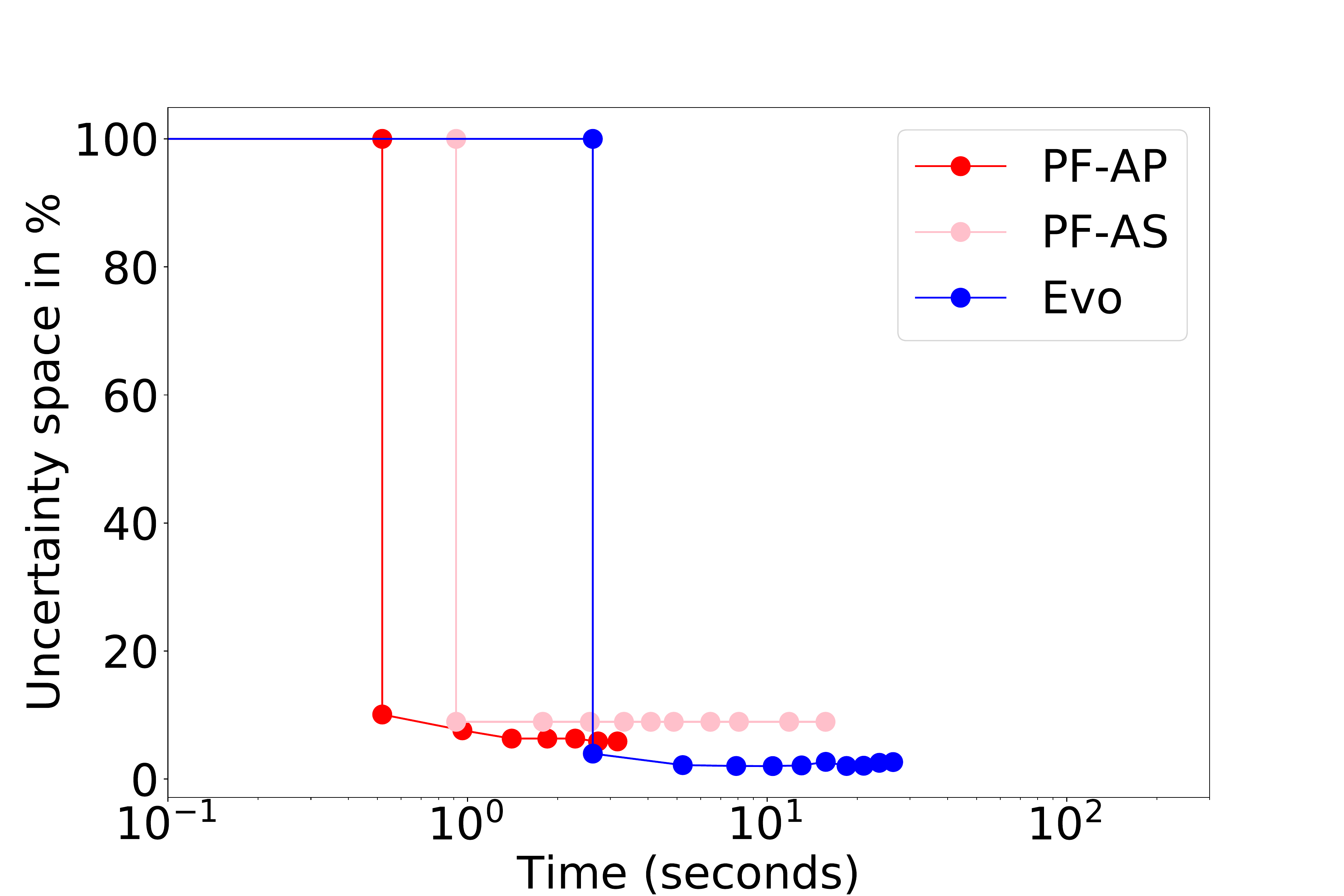}}

		&	
		\subfigure[\small{Frontier of Evo (job 9, 2d)}]
		{\label{fig:batch-9-3-evo}\includegraphics[height=3.2cm,width=5.5cm]{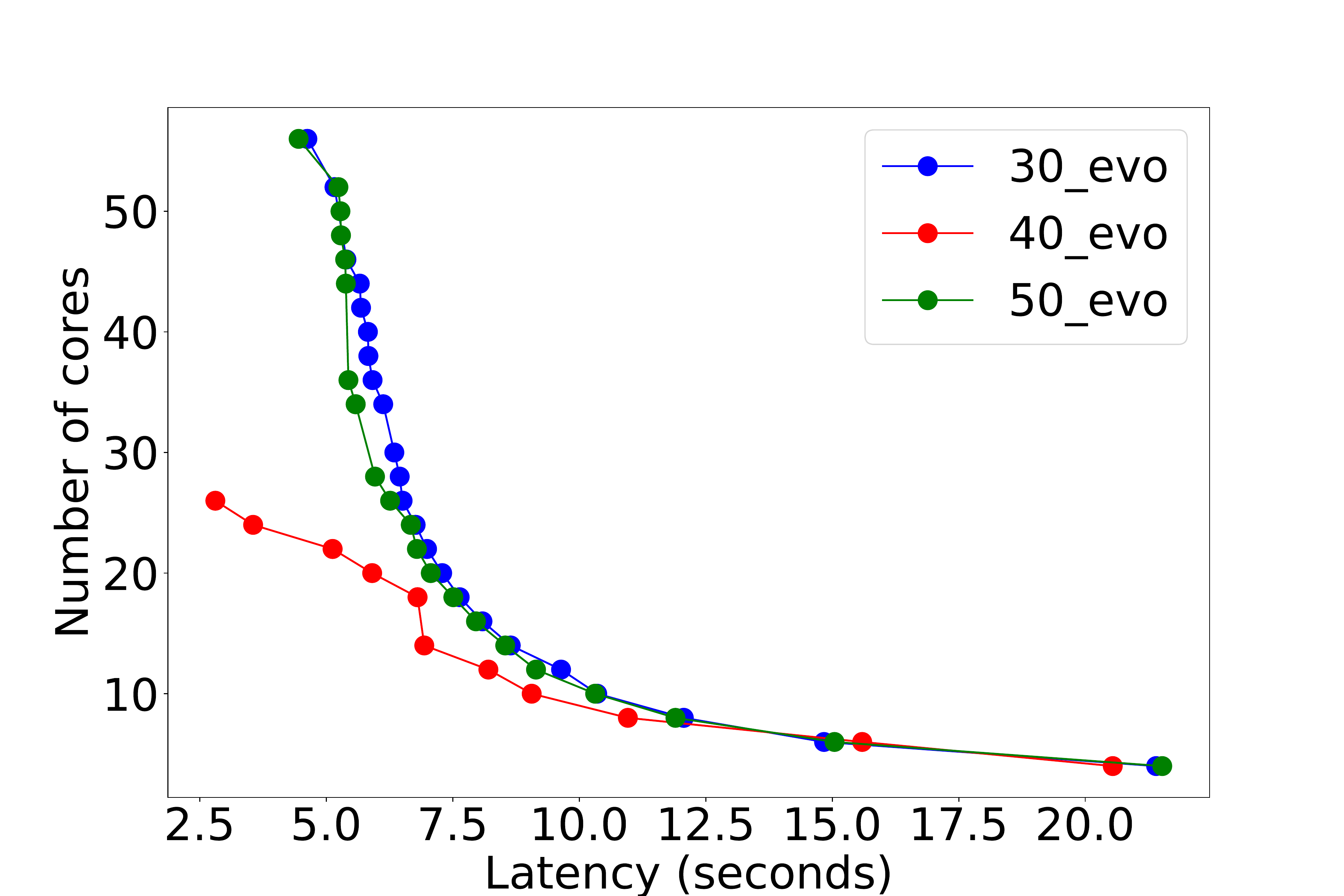}}
	
		&	
		\subfigure[\small{Uncertain space of all 258 jobs}]
		{\label{fig:batch-all-2sec}\includegraphics[height=3.2cm,width=5.5cm]{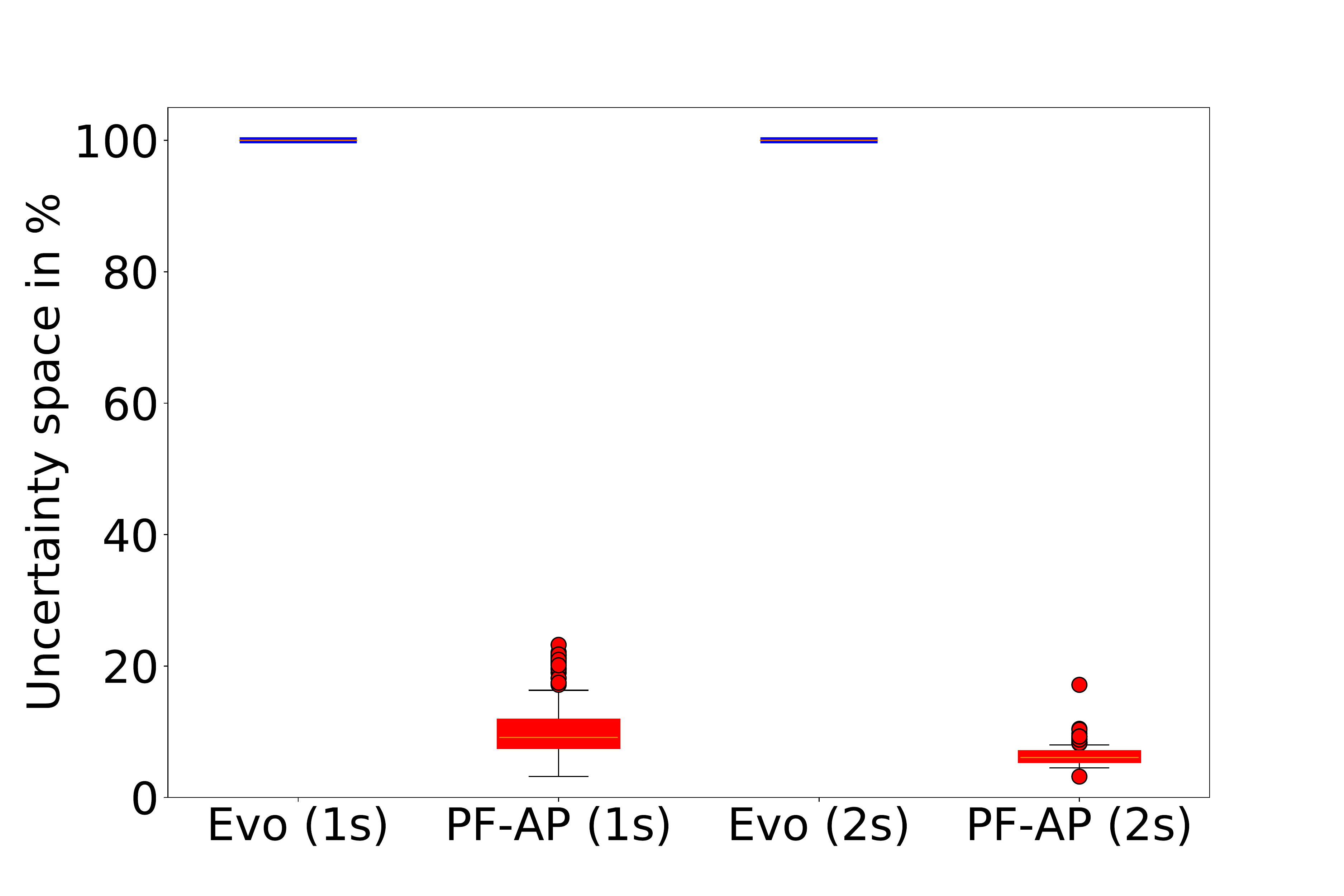}}



%
%
%

%
%
%
	
	\end{tabular}
\vspace{-0.2in}
\caption{\small Comparative results on multi-objective optimization using 258 batch workloads}
 \label{fig:moo}
\vspace{-0.1in}
\end{figure*}

In this section, we compare our MOO approach to popular MOO techniques and perform an end-to-end comparison to the state-of-the-art OtterTune~\cite{VanAken:2017:ADM} system.


\textbf{System.}
We chose Apache Spark as the data analytics system because it allows us to run analytics including both SQL queries and ML tasks. Within our dataflow optimizer, the modeling engine employs two modeling tools, GP models from OtterTune~\cite{VanAken:2017:ADM} and our custom deep neural network (DNN) models. 
Since DNN models are more complex than GP models, they increase computational cost in MOO. 
Therefore, we use DNN models as the default when we evaluate efficiency of  MOO methods, and will switch to GP models when we conduct end-to-end comparison to OtterTune.
Our modeling engine is implemented using a mix of PyTorch, Keras and Tensorflow to implement and train different models. 
The trained models interface with our MOO module through network sockets.


Our MOO module is implemented in Java and invokes a solver for constrained  optimization (CO). Our system supports several solvers including our MO-GD solver (\S\ref{subsec:dnn}) and the  Knitro~\cite{Knitro} solver. To solve a single CO problem,  Knitro with 16 threads takes 17 and 42 minutes to run on GP and DNN models, respectively. In contrast, MO-GD  with 16 threads takes 0.1-0.5 second while achieving the same or lower value of the target objective. 
We also built an approximate solver based on CPLEX~\cite{Cplex} by relaxing our CO problem to continuous non-linear programming. However, it exhibited a scalability issue in handling DNN models.  Therefore, we use MO-GD as the default solver for our experiments. 


\textbf{Workloads.}
We used two benchmarks for evaluation. 

{\em Batch Workloads} (TPCx-BB): 
We used the TPCx-BB (BigBench) benchmark~\cite{TPCx-BB} designed to model mixed workloads in big data analytics, with a scale factor 100G. TPCx-BB includes 30 templates, including 14 SQL queries, 11 SQL with UDF, and 5 ML tasks, which we
  modified to run on Spark. 
We parameterized the 30 templates to create 258 workloads and ran them under different configurations, totaling 19528 traces,  each with 360 runtime metrics (OtterTune~\cite{VanAken:2017:ADM} suggests similar numbers of traces for training). These traces were then used to train workload-specific models for latency, CPU utilization, IO cost, etc.  After hyperparameter tuning, our latency model has 4 hidden layers, each with 128 nodes, and uses  ReLU 
as the activation function.
Adaptive moment estimation (Adam)~\cite{kingma2014adam} was used to run backpropogation with learning rate = 0.1, weight decay = 0.1, max\_iter = 100, and early stop patience = 20.
OtterTune~\cite{VanAken:2017:ADM} suggests using feature selection to focus on the most important ($\sim$10) parameters for tuning; hence,  we ran  MOO  over the most important 12 parameters  of Spark, including parallelism,  the number of executors, the number of core per executor, memory per executor, shuffle compress, etc. 

{\em Streaming Workloads}: 
We also created a streaming benchmark by extending a prior study~\cite{LiDS15} on click stream analysis, 
including  5 SQL  templates with UDFs and 1 ML template. 
We then created 63 workloads from templates using parameterization, and collected traces for training workload models for latency and throughput.

{\bf Hardware}.
Our system was deployed on a cluster with one gateway and 20 compute nodes. The compute nodes are CentOS based with 2xIntel Xeon Gold 6130 processors and 16 cores each, 768GB of memory, and RAID disks.\cut{The cluster also contains two 1Gb/sec and 100Gb/sec networks for management and data transfer, respectively.} 
Each MOO algorithm  runs on a dedicated compute node, potentially with multi-threading.

%% file: moo_experiments.tex
\subsection{Comparison to MOO Methods}
\label{subsec:moo}

We first compare our Progressive Frontier (PF) algorithms, \sdnn\ and \pdnn, to three major MOO methods, Weighted Sum (WS)~\cite{marler2004survey}, Normalized Constraints (NC)~\cite{messac2003nc}, and NSGA-II~\cite{Deb:2002:FEM} from JMetal~\cite{JMetal} in  the family of Evolutionary (Evo) methods~\cite{Emmerich:2018:TMO}. (As~\cite{Emmerich:2018:TMO} points out, more recent Evo methods concern {\em many} objective optimization and preference modeling, hence orthogonal to our current work.)
For each algorithm, we request it to generate increasingly more Pareto points (10, 20, 30, 40, 50, 100, 150, 200), which are called probes, as more computing time is invested. 

\textbf{Expt 1: Batch 2D.}
We start with the batch workloads where the objectives are latency  and cost (simulated by the number of cores used). As results across different jobs are consistent, we first show details using job 9.
To compare \sdnn\ and \pdnn\ to WS and NC, Fig.~\ref{fig:batch-9-3-time} shows the uncertain space  when  more Pareto points are requested over time. Here, the uncertain space is the percentage of the total objective space that the algorithm is uncertain about (Def.~\ref{def:uncertainspace}). 
Initially at 100\%, it starts to reduce when the first Pareto set (of up to 10 points) is generated.
A main observation that WS and NC take long to run, e.g., about 47 seconds to generate the first Pareto set. Such delay means that they are not suitable for recommending job configurations under stringent time constraints. 
In comparison, our PF approximate sequential (\sdnn) and parallel (\pdnn)  algorithms reduce uncertain space much more quickly, e.g., with the first Pareto set generated under 1 second by \pdnn. 
\sdnn\ does not work as well as \pdnn\ because \sdnn\ is a sequential algorithm, and hence the quality of Pareto points found in the early stage have a severe impact on the later procedure. Given that our algorithm is approximate, it could happen that one low-quality result in the early stage leads to overall low-quality Pareto frontier. In contrast, the \pdnn\ is a parallel algorithm and hence one low-quality result won't have as much  impact as in \sdnn.

Fig.~\ref{fig:batch-9-3-ws-nc}  shows the Pareto frontiers of WS and NC generated after 47 seconds, where 
the Utopia (hypothetically optimal) point is at the lower left corner.
WS is shown to have {\em poor coverage} of the Pareto frontier, e.g., returning only 3 points although 10 were requested. 
NC generates more points (8) on the Pareto frontier. However, it still provides fewer points and less information than the Pareto frontier of  \pdnn\ (12 points) shown in Fig.~\ref{fig:batch-9-3-pf}, constructed using only 3.2 seconds. 
These frontiers also show that  latency and cost do compete for resources, hence with tradeoffs. 

We next compare \pdnn\ to Evo, as Fig.~\ref{fig:batch-9-3-time-evo} shows. Although Evo runs faster than WS and NC, it still fails to generate the first Pareto set until after 2.6 seconds. As noted before, such a delay is still quite high for our target use case of serverless databases. Fig.~\ref{fig:batch-9-3-evo} shows another issue of Evo: the Pareto frontiers generated over time are not consistent. For example, the frontier generated with 30 probes indicates that if one aims at latency of 6 seconds, the cost is around 36 units. The frontier produced with 40 probes shows the cost to be as low as 20, while the frontier  with 50 probes changes the cost to 28. Recommending configurations based on such inconsistent information is highly undesirable. 

Finally, Fig.~\ref{fig:batch-all-2sec} compares \pdnn\ against Evo for {\bf all 258} jobs, when we impose a 1-second (or 2 second) constraint for making a configuration recommendation to balance  cost and latency, e.g.,  when a serverless database needs to be started. Evo fails to generate any Pareto set under 1 seconds (with 100\% uncertain space) and hence cannot make any recommendation. In contrast, our \pdnn\ can generate Pareto sets under 1 seconds for all 258 jobs, with a median of 9.2\% uncertain space across all jobs. For a 2 second time constraint, Evo still can not make any recommendation. Our \pdnn\ can generate Pareto sets under 2 seconds for all 258 jobs, with a median of 6.1\% uncertain space  across all jobs. 

\begin{figure*}[t]
	\centering
	\hspace{-6cm}
	
	\begin{tabular}{lcc}
		
		\subfigure[\small{Uncertain space (job 54, 3d)}]
		{\label{fig:stream-54-time-3d}\includegraphics[height=3.2cm,width=5.5cm]{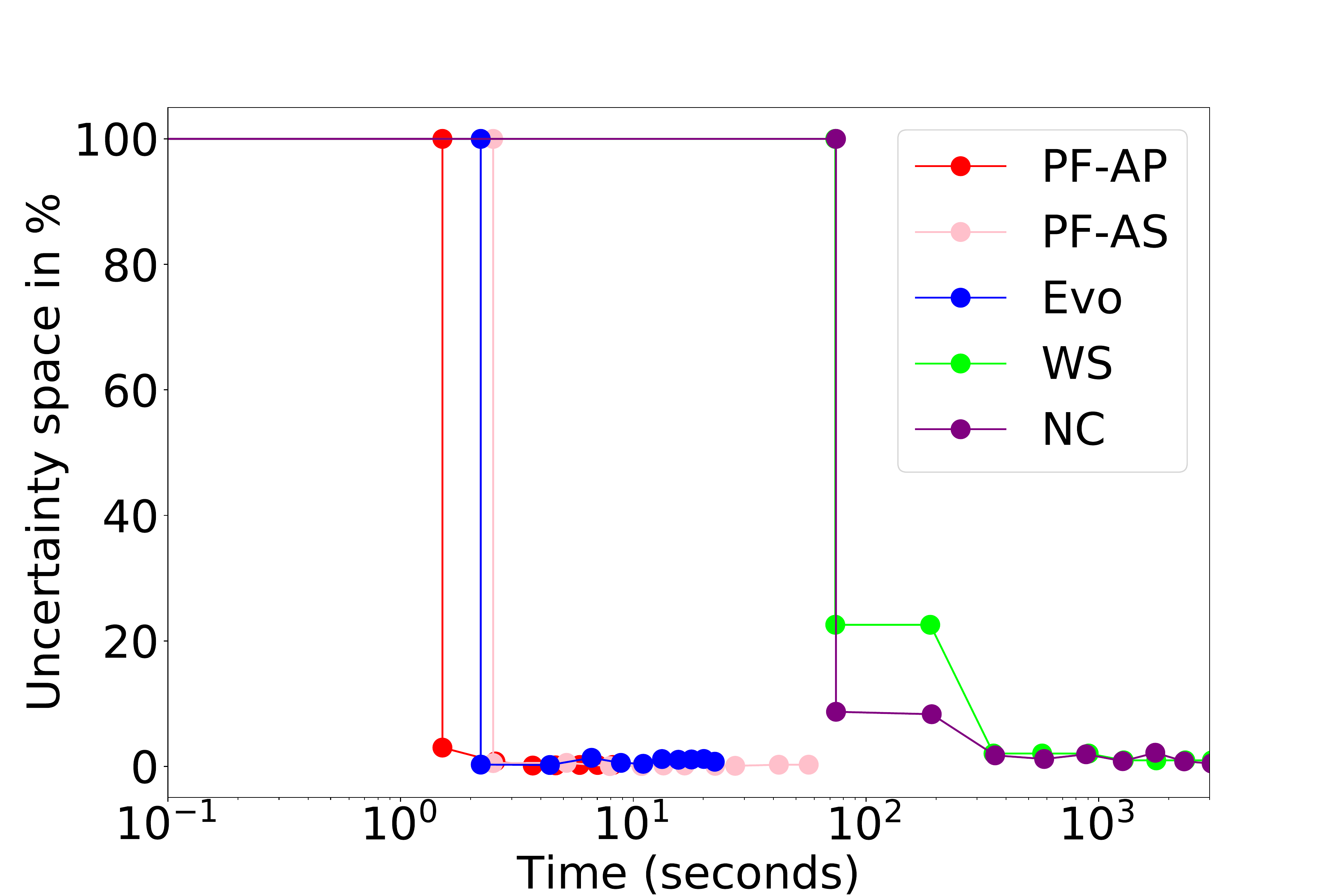}}

		&	
		\subfigure[\small{Frontier of WS (job 54, 3d)}]
		{\label{fig:stream-54-ws-3d}\includegraphics[height=3.2cm,width=5.5cm]{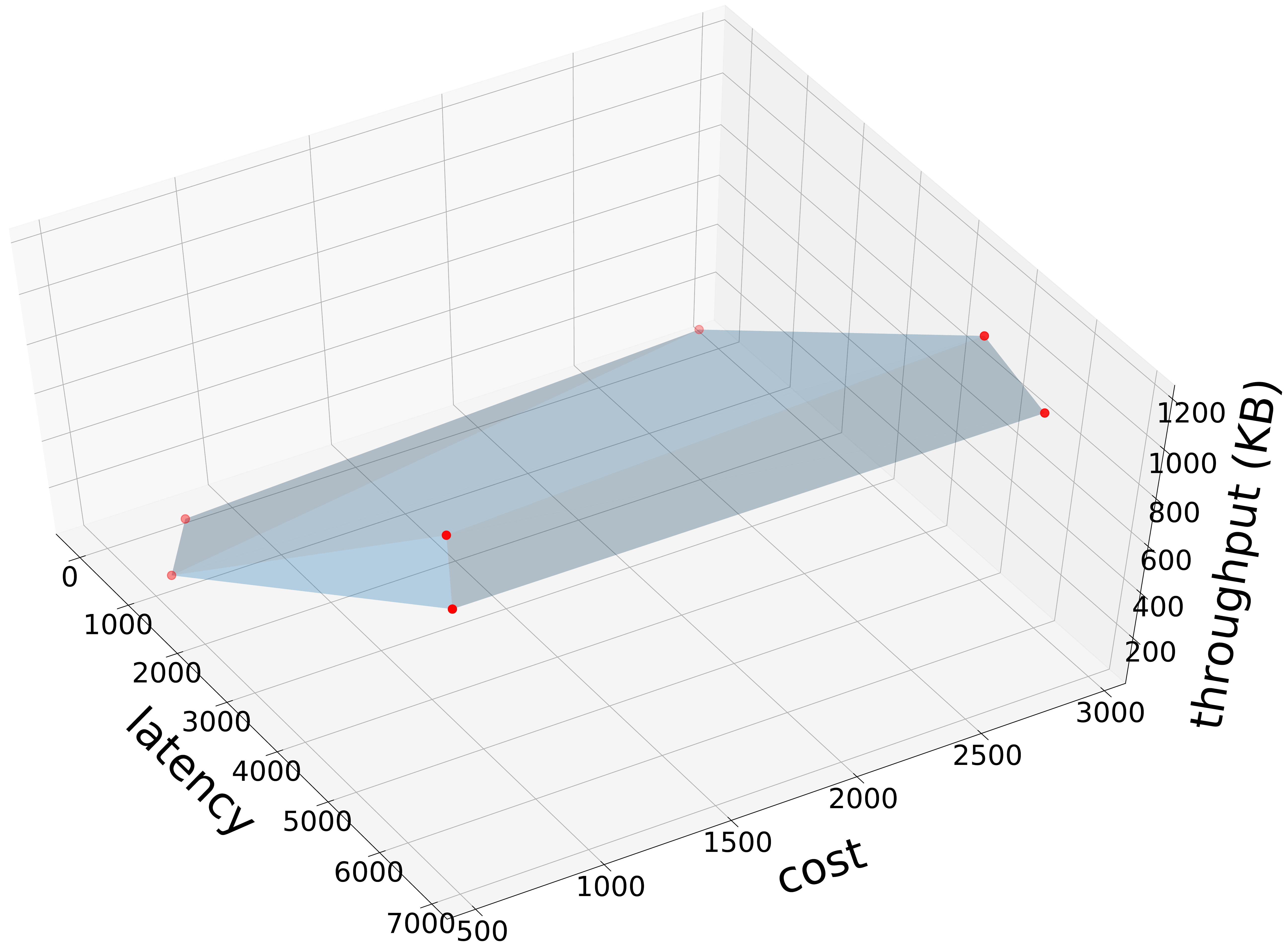}}

		&
		\subfigure[\small{Frontier of NC (job 54, 3d)}]
		{\label{fig:stream-54-nc-3d}\includegraphics[height=3.2cm,width=5.5cm]{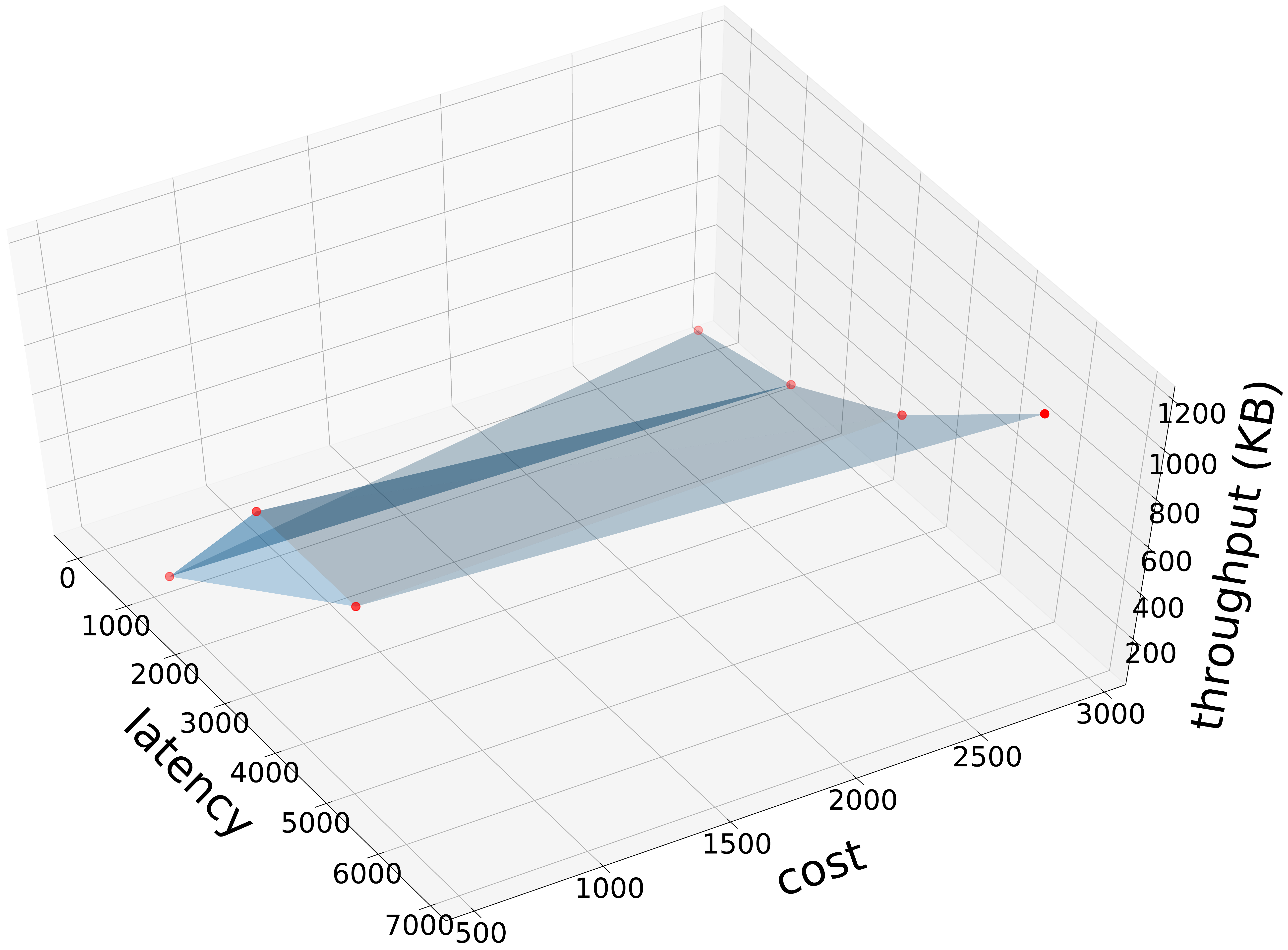}}

		\\	
		
		\subfigure[\small{Frontier of PF (job 54, 3d)}]
		{\label{fig:stream-54-pf}\includegraphics[height=3.2cm,width=5.5cm]{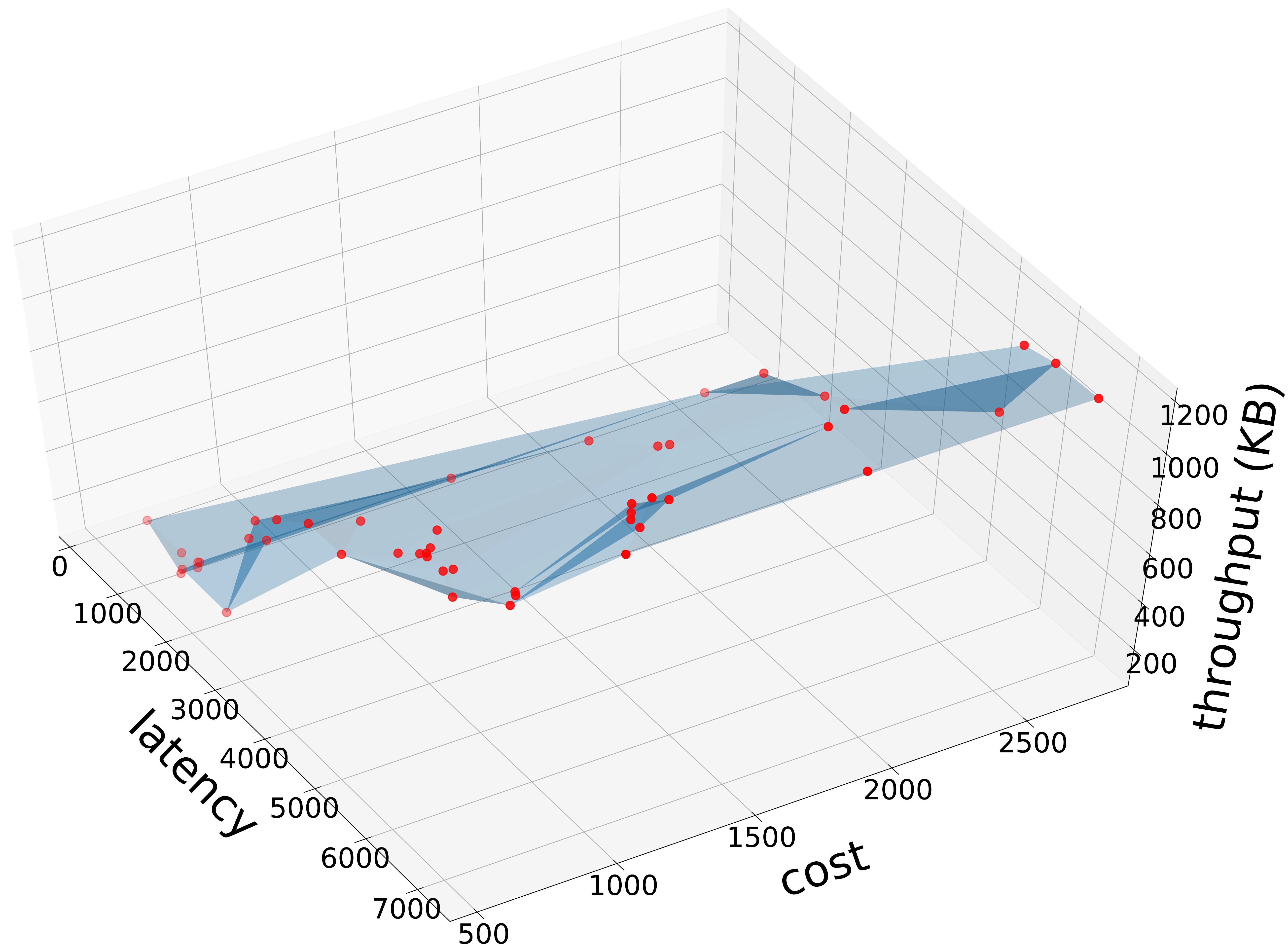}}
	
		&
		\subfigure[\small{Uncertain space under 1 or 2 seconds (2D)}]
		{\label{fig:stream-all-2sec-2d}\includegraphics[height=3.2cm,width=5.5cm]{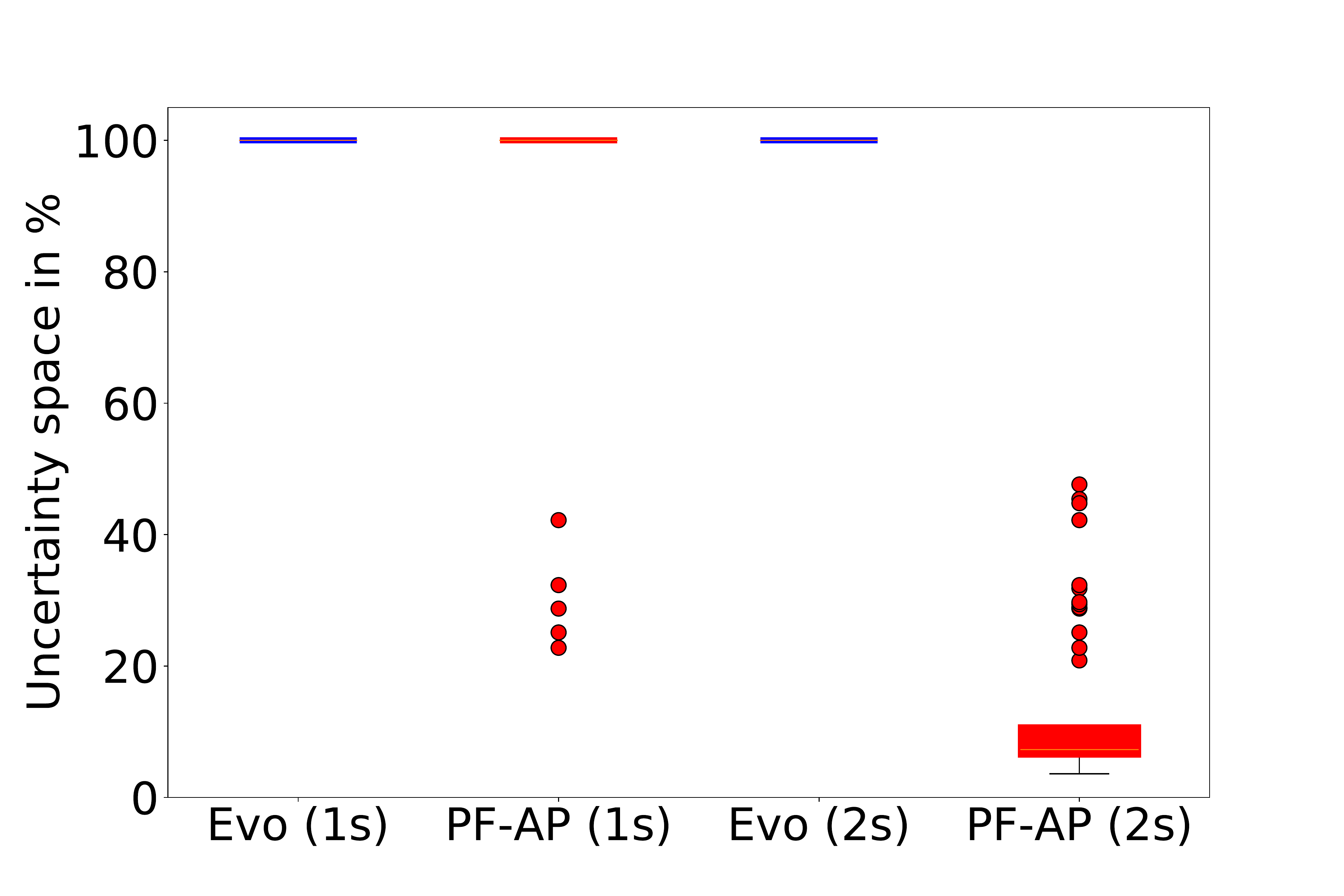}}
		
		&	
		\subfigure[\small{Uncertain space under 1 or 2.5 seconds (3D)}]
		{\label{fig:stream-all-2sec-3d}\includegraphics[height=3.2cm,width=5.5cm]{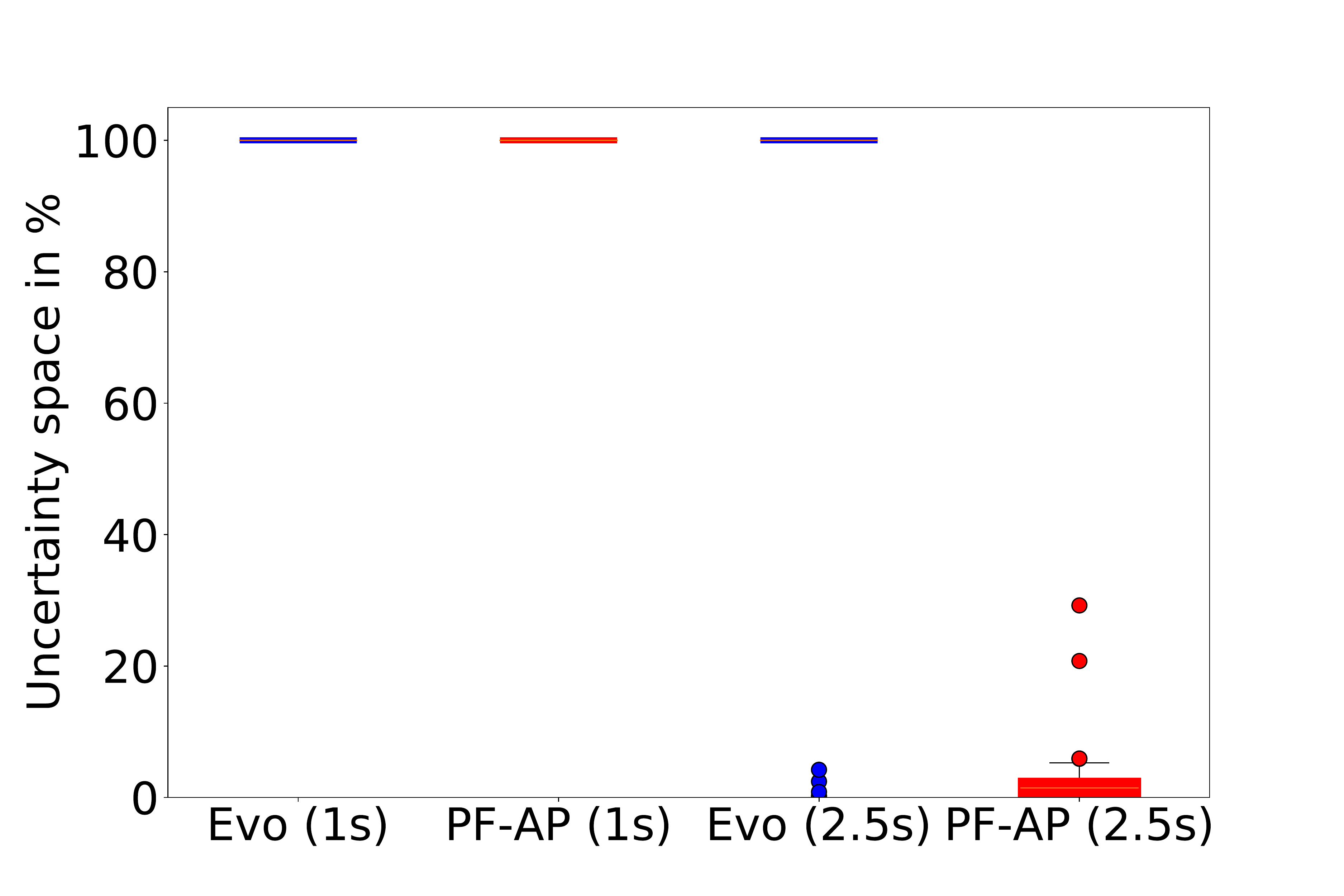}}
		
	\end{tabular}
	\vspace{-0.2in}
	\caption{\small Comparative results on multi-objective optimization using 63 streaming workloads}
	\label{fig:moo_3d}
	\vspace{-0.1in}
\end{figure*}

\textbf{Expt 2: Streaming 2D and 3D.} We next use the streaming workload under 2 objectives, where the objectives are average latency (of output records) and throughput (the number of records per second), as well as under 3 objectives, where we add (simulated) cost as the 3rd objective.
As the results for different jobs are similar, we illustrate them using job 54 under 3D, while additional results are available in \techreport{Appendix~\ref{appendix:results}}{our technical report~\cite{Song2020}}.
Fig.~\ref{fig:stream-54-time-3d} confirms that WS and NC take long, e.g., 74 and 75 seconds, respectively, to return the first Pareto set, where our \pdnn\ computes the first Pareto set with 1.5 seconds. 
Evo method  again fails to produce the first Pareto set under 2.2 seconds. 
Fig.~\ref{fig:stream-54-ws-3d} and Fig.~\ref{fig:stream-54-nc-3d} show that WS and NC again have poor coverage of the frontier (7 points only for each), while Fig.~\ref{fig:stream-54-pf} shows that PF can offer much better coverage of the frontier using less time. 
Again, we observe that Evo returns inconsistent Pareto frontiers as more probes are made, whose plots are left to \techreport{Appendix~\ref{appendix:results}}{our technical report~\cite{Song2020}} due to space limitations.

Fig.~\ref{fig:stream-all-2sec-2d} and ~\ref{fig:stream-all-2sec-3d} summarize the running time of our \pdnn\ and Evo for 2D and 3D cases. 
For all 2D  jobs, Evo fail to  meet the 1-second or 2-second constraint. 
Our \pdnn\ can generate Pareto sets for 5 jobs under 1 second, and all 63 jobs under 2 seconds, with a median of 8\% uncertain space.  
For 3D jobs, we give a slightly looser constraint, 2.5 seconds, to accommodate the increased dimensionality. \pdnn\ can generate Pareto sets for all 63 jobs under 2.5 seconds, with  a median of 2\%  uncertain space, whereas Evo can generate Pareto sets for 5 out of 63 under 2.5 seconds. 

%

\textbf{Summary.} Across all 321 workloads tested, WS and NC fail to generate Pareto sets under 40 seconds. Evo is faster but still fails to generate the first Pareto set for most jobs under the constraints of 2 - 2.5 seconds. In contrast,  our \pdnn\ algorithm takes less than 2.5 seconds to generate a decent Pareto set for all jobs, offering 2.6x to 50x speedup over other methods in achieving so.\cut{Hence, \pdnn\ is more suitable for recommending job configurations under stringent time constraints (e.g., when starting a new database instance), while offering more accurate information later for auto-scaling  or running a recurring job. }
In addition, \pdnn\ overcomes the poor coverage (WS and NC) and inconsistency (Evo) issues that other MOO algorithms have.  

\begin{figure*}[t]
	\centering
	\hspace{-6cm}
	
	\begin{tabular}{lcc}

		\subfigure[\small{Batch (0.5,0.5), accurate models}]
		{\label{fig:batch-ot-5}\includegraphics[height=3.2cm,width=5.5cm]{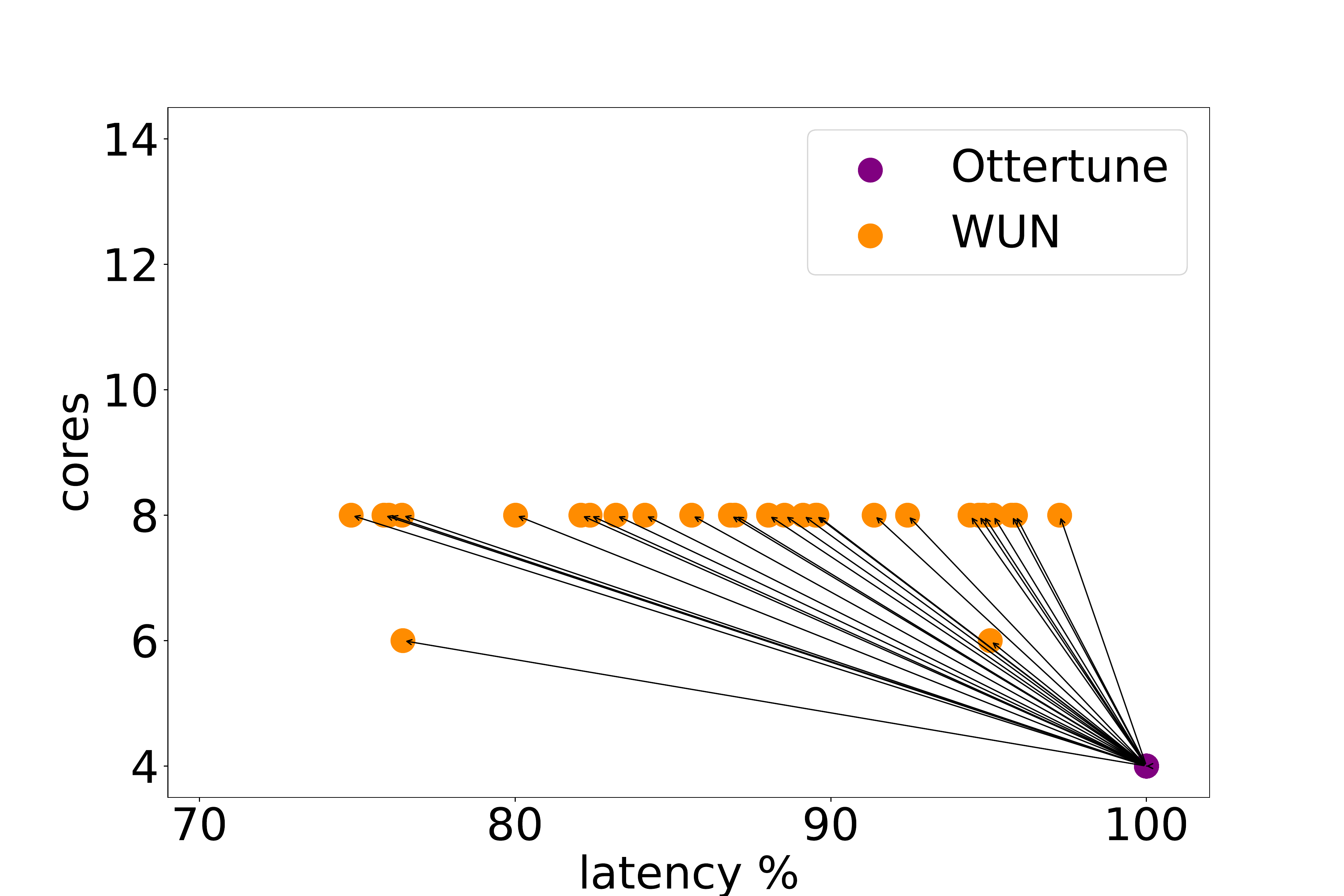}}

		&	
		\subfigure[\small{Batch (0.9,0.1), accurate models }]
		{\label{fig:batch-ot-9}\includegraphics[height=3.2cm,width=5.5cm]{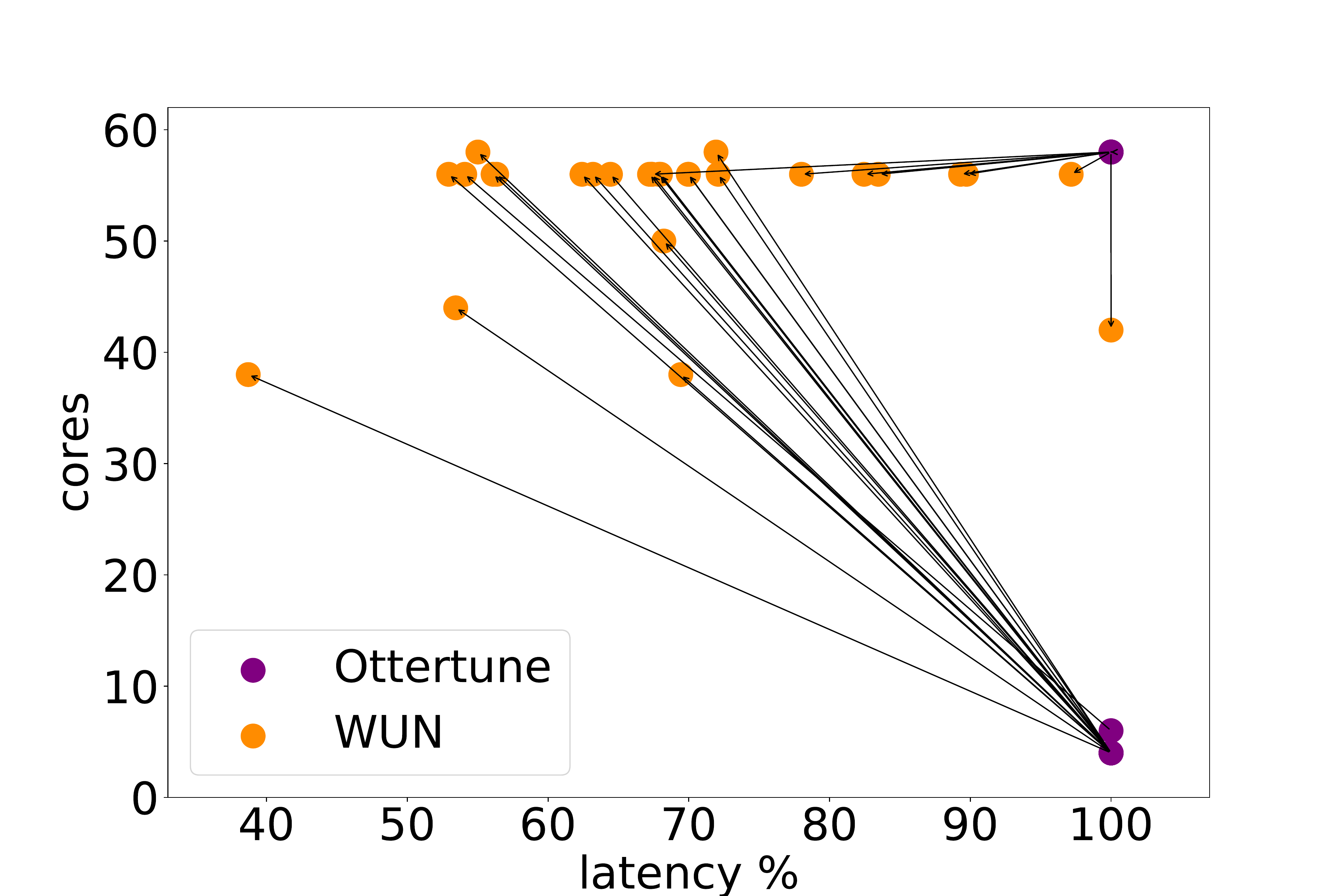}}

		&
		\subfigure[\small{Stream (0.5,0.5), accurate models }]
		{\label{fig:stream-ot-5}\includegraphics[height=3.2cm,width=5.5cm]{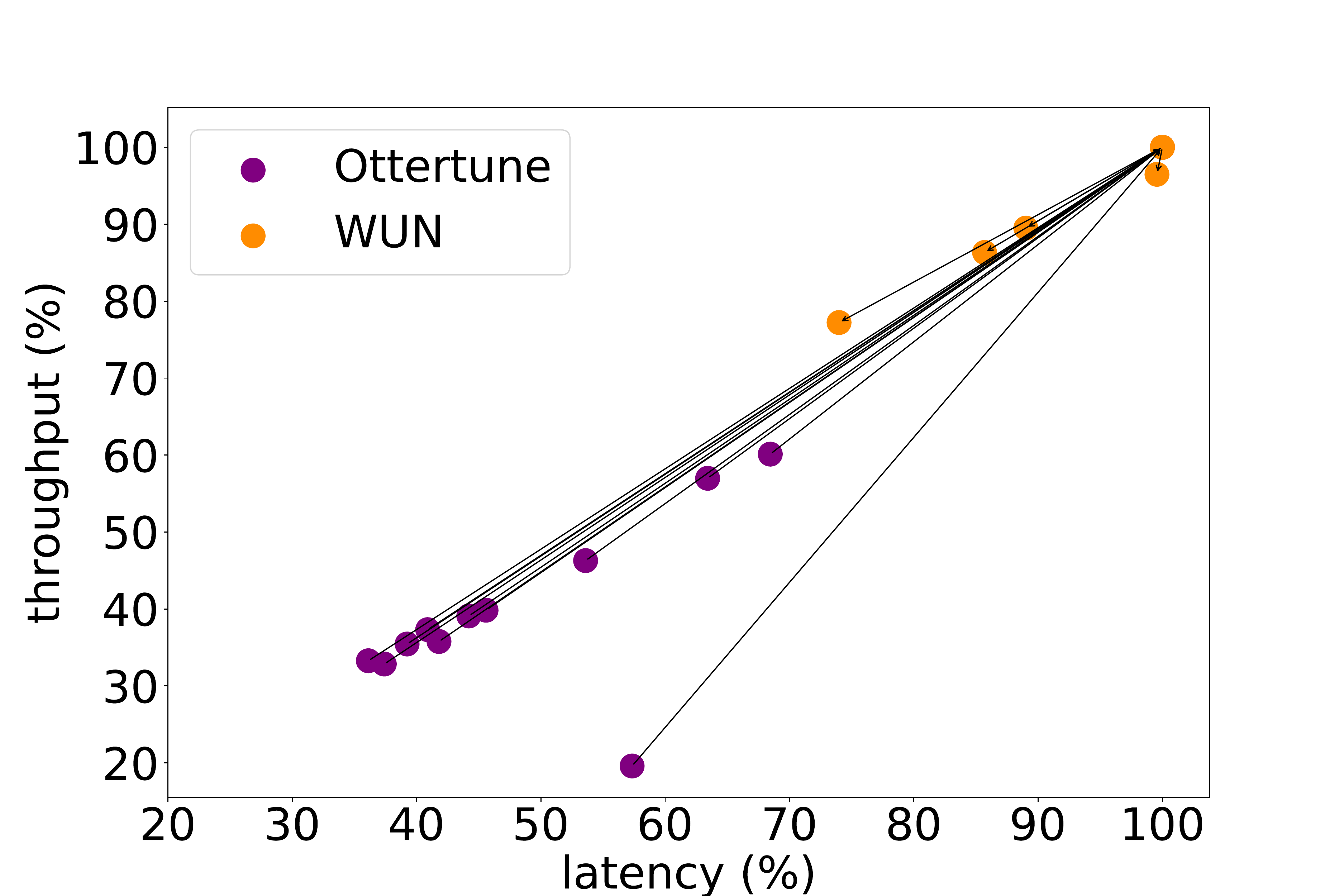}}

		\\	
				
		\subfigure[\small{Stream (0.9,0.1), accurate models }]
		{\label{fig:stream-ot-9}\includegraphics[height=3.2cm,width=5.5cm]{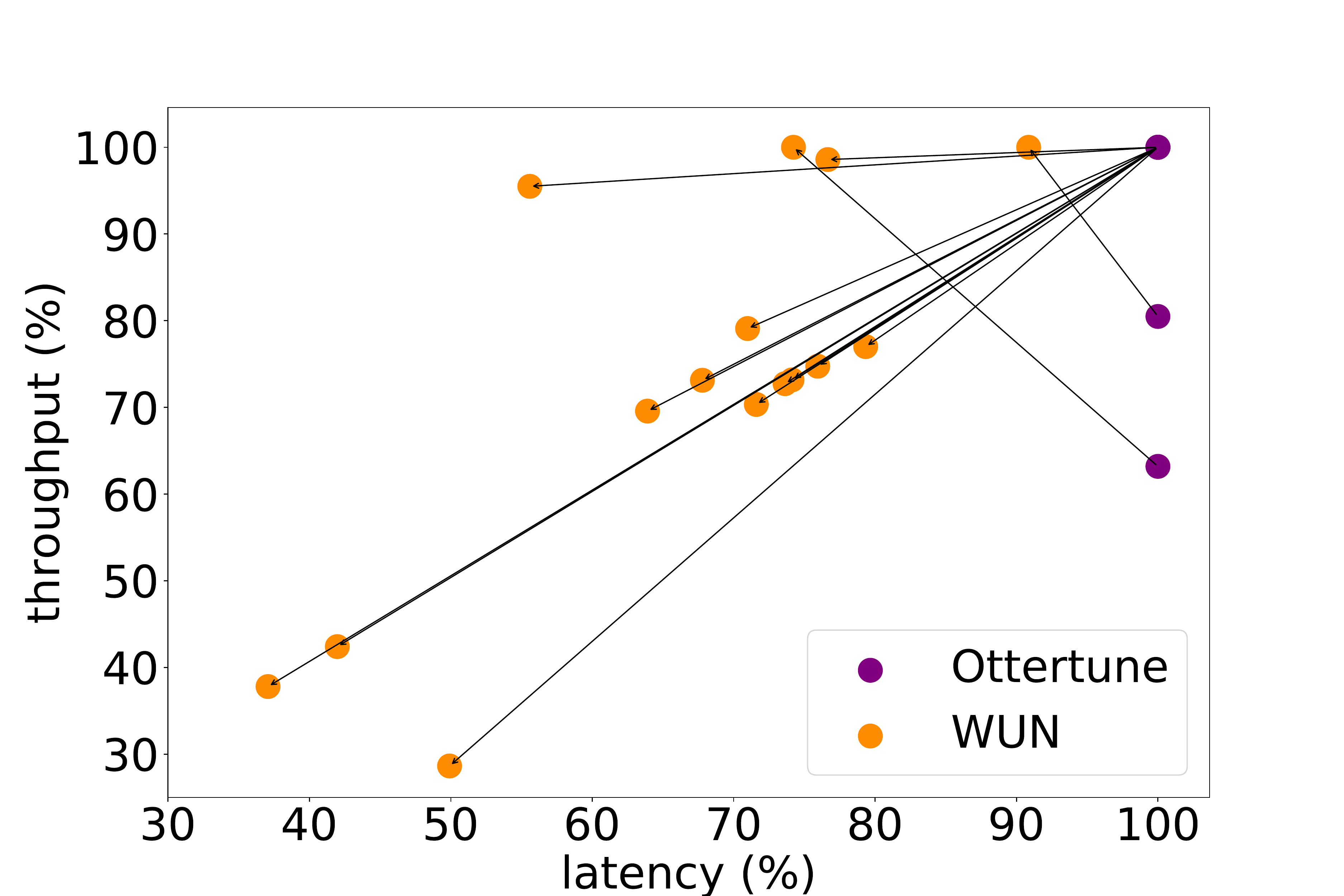}}

		&
		\subfigure[\small{Batch (0.5,0.5),  inaccurate models}]
		{\label{fig:batch-5-inaccr-2d}\includegraphics[height=3.2cm,width=5.5cm]{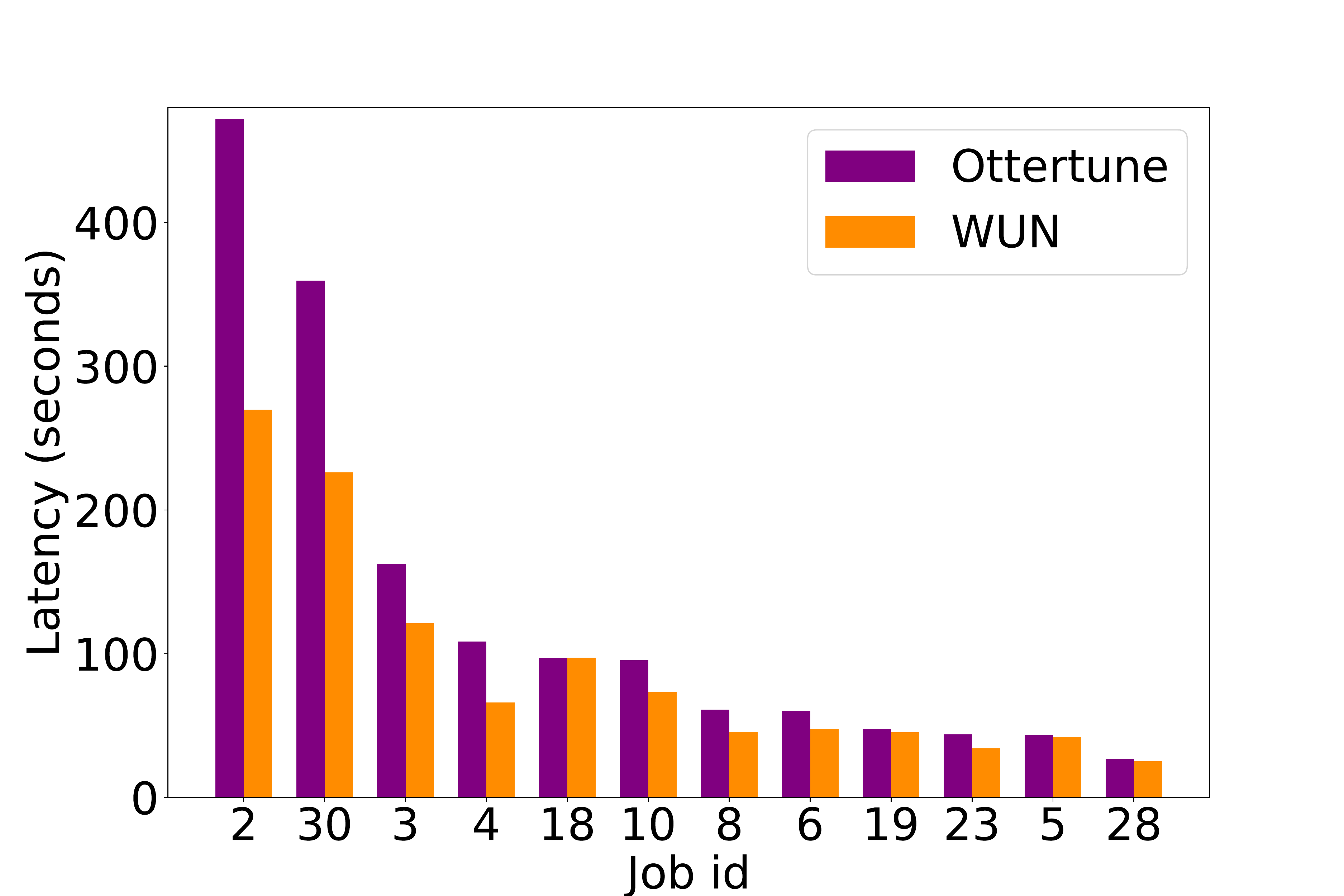}}
		
		&
		\subfigure[\small{Batch (0.9,0.1), inaccurate models}]
		{\label{fig:batch-9-inaccr-2d}\includegraphics[height=3.2cm,width=5.5cm]{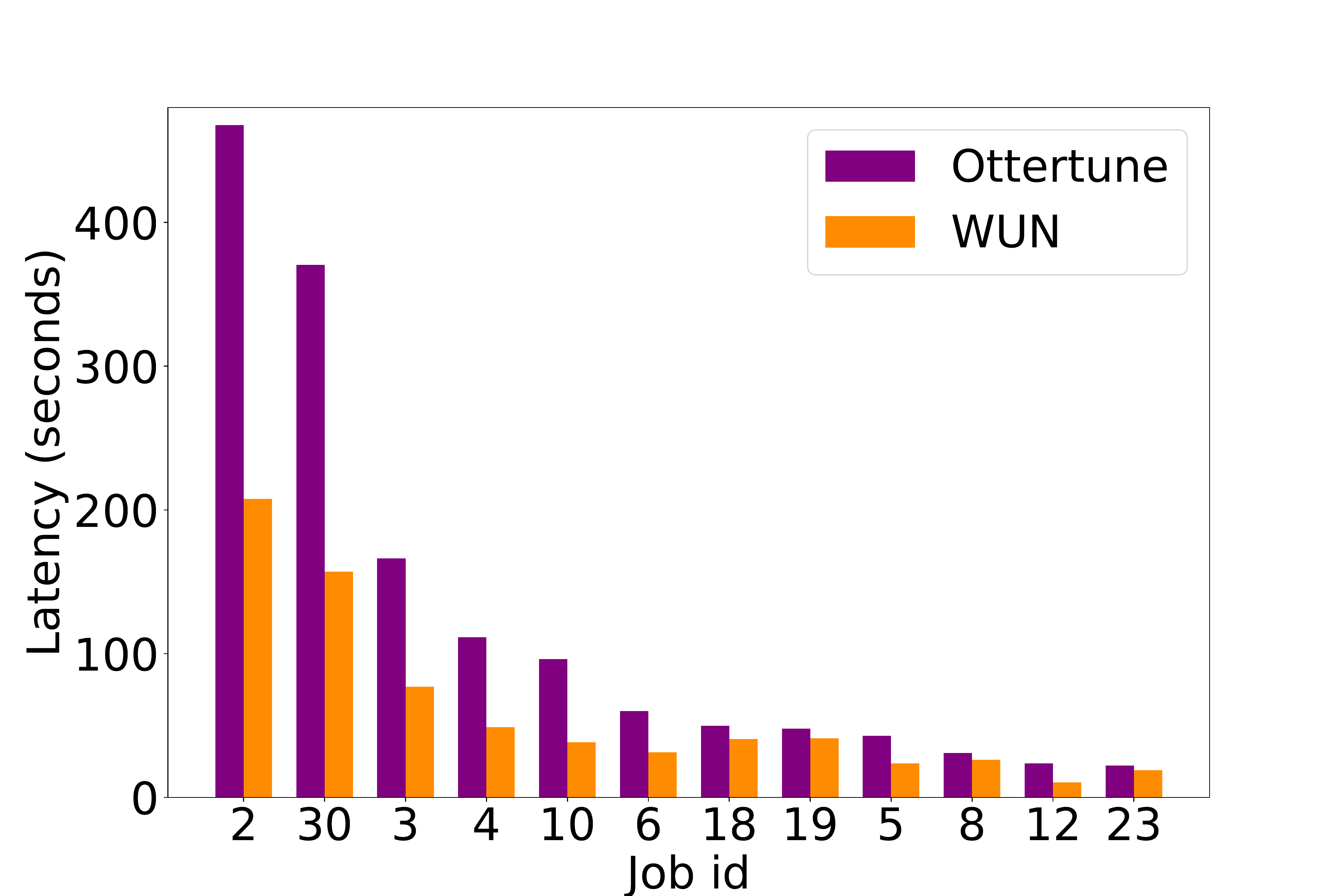}}

%


		
		
	\end{tabular}
	\vspace{-0.2in}
	\caption{\small Comparative results to Ottertune on single-objective and multi-objective optimization }
	\label{fig:moo_3d}
	\vspace{-0.1in}
\end{figure*}

\subsection{End-to-End Comparison}
Next we perform an end-to-end comparison of the configurations recommended by our MOO  against  those  by Ottertune ~\cite{VanAken:2017:ADM},  a state-of-the-art automatic tuning system as described in \S\ref{subsec:priorwork}.
\cut{Our system differs from Ottertune in two aspects:
(1)~We can use a variety of models while Ottertune uses only GP models.
(2)~We support MOO while Ottertune performs single-objective  optimization. }In this study, since we consider the effect of models, we follow the standard practice in machine learning to separate workloads into training and test sets: For TPCx-BB, we selected a random sample of 30 workloads, one from each  template, as the test workloads. We use the trace data of other workloads as historical data to train predictive models for the test workloads. We did it similarly for the stream benchmark, with a sample of 15 test workloads.

\textbf{Expt 3: Accurate models.}
First, we assume that the learned models are accurate and treat the model-predicted value of an objective as the true value, for any given configuration. To ensure fair comparison, we use the GP models from Ottertune in both systems. For each of the following 2D workloads, our system runs the PF algorithm to compute the Pareto set for each workload and then  the Weighted Utopia Nearest (WUN) strategy (\S\ref{subsec:recommendation})  to recommend a configuration from the set.  
As Ottertune supports only single-objective (SO) optimization, we apply a weighted approach~\cite{Zhang:2019:EAC} that  combines $k$ objective functions into a single objective,  $\sum_{i=1}^k w_i \Psi_i(\bm{x})$, with $\sum_i w_i = 1$, and then call Ottertune to solve a SO problem. The weight vector, $\bm{w}$, is a hyperparameter that has to be set before running Ottertune for a workload. 

It is important to note that the weights are used very differently in these two systems:
\wun\ applies $\bm{w}$  in the objective space  to select one  configuration from its Pareto set. 
In contrast, the weighted method for Ottertune applies $\bm{w}$ to construct a SO problem and hence obtains a single solution. It is known from the theory of Weighted Sum~\cite{marler2004survey} that 
even if one tries many different values of $\bm{w}$, the weighted method cannot find many diverse Pareto points, which is confirmed by the sparsity of the Pareto set in  Fig.~\ref{fig:stream-54-ws-3d} where Weight Sum tried different $\bm{w}$ values.

{\em Batch 2D}. For the 2D batch workloads,  
We compare the recommendations of  \wun\ to those of Ottertune.  Fig.~\ref{fig:batch-ot-5} shows the comparison when $\bm{w}$=(0.5, 0.5), indicating that the application wants balanced results between latency and cost. Since TPCx-BB workloads have 2 orders of magnitude difference in latency, we normalize the latency  of each workload (x-axis) by treating the slower one between \wun\ and Ottertune as 100\% and the faster as a value less than 100\%. The number of cores (y-axis) allowed in this experiment is [4, 58].
For all 30 workloads, Ottertune recommends the smallest number of cores (4),  favoring small numbers of cores at the cost of high latency. \wun\ is adaptive to the application requirement of balanced  objectives, using 2-4 more cores for each workload to enable up to 26\% reduction of latency. 
When we change the weights to (0.1, 0.9), favoring cost to latency, both systems recommend using 4 cores; the plot is omitted in the interest of space.
Fig.~\ref{fig:batch-ot-9} shows the comparison when $\bm{w}$=(0.9, 0.1), indicating strong application preference for low latency. For 19 out of 30 workloads, Ottertune still  recommends 4 cores because that is the solution returned even using the 0.9 weight for latency. In contrast, \wun\ is more adaptive, achieving lower latency than Ottertune for all 30 workloads with up to 61\% reduction of latency. In addition, for 8 out of 30 workloads, \wun\ dominates Ottertune in both objectives, saving up to 33\% of latency while using fewer cores -- in this case, Ottertune's solution is not Pareto optimal.



{\em Streaming 2D}. For 15 streaming workloads, 
Fig.~\ref{fig:stream-ot-5} shows the comparison when $\bm{w}$=(0.5, 0.5), where we normalize both latency (x-axis) and throughput (y-axis) using the larger value between \wun\ and Ottertune. 
The two systems mainly indicate tradeoffs: Ottertune recommends low-latency, low-throughput configurations, while \wun\ recommends high-latency, high-throughput configurations. 
When the application changes the preference to $\bm{w}$=(0.9, 0.1), again \wun\ is more adaptive, achieving lower latency for all 15 workloads with up to 63\% reduction of latency. For 2 out of 15 jobs, \wun\ dominates Ottertune in both objectives, e.g., 36\% reduction of latency and 30\% increase of throughput. Again, Ottertune's solution is not Pareto optimal here.


\textbf{Expt 4: Inaccurate models.}
We next consider a realistic setting where learned models are not accurate. In this case, our MO-GD solver uses the variance of the model prediction for a given objective to obtain a more conservative estimate during optimization. In addition, we obtained a more accurate latency model for the 30 TPCx-BB workloads using our customized DNN implementation. Note that our optimizer does not favor any particular model. Our experiment simply uses the DNN latency model to demonstrate the flexibility of our optimizer, while Ottertune can only use its GP model. 
By simply changing the model, we reduce latency by 5\% over Ottertune when solving 1D optimization on latency. 

Next we consider 2D optimization over latency and cost. For $\bm{w}$=(0.5, 0.5), we take recommendations from both systems and measure the actual latency and  cost (num. of cores times latency) on our cluster. Fig.~\ref{fig:batch-5-inaccr-2d} shows detailed latency results of top 12 long-running workloads. Since both systems use low numbers of cores, the cost plot is quite similar to the latency plot, hence omitted here. Most importantly, to run the full TPCx-BB benchmark, our system outperforms Ottertune with 26\% savings on running time while using 3\% less cost. 
For $\bm{w}$=(0.9, 0.1), Ottertune's recommendations vary from those for (0.5, 0.5) with only 6\% reduction of total running time, while our recommendations lead to 35\% reduction. As Fig.~\ref{fig:batch-9-inaccr-2d} shows, our system outperforms Ottertune by 49\% reduction of total running time, with 48\% increase of cost, which matches the application's strong preference for latency. 


%% file: related_work.tex
\section{Related Work}
\noindent
\textbf{Multi-objective optimization}  
for SQL~\cite{Ganguly:1992:QOP,Kllapi:2011:SOD,Trummer:2014:ASM,TrummerK15} enumerates a {\em finite} set of query plans 
and selects  Pareto-optimal ones based  on cost---in contrast to our need for searching through an
an {\em infinite} parameter space.
MOO approaches for workflow scheduling ~\cite{Kllapi:2011:SOD} differ from our MOO in the parameter space and solution.

\noindent
\textbf{Resource management.}
TEMPO~\cite{tan2016tempo} addresses resource management for SQL databases in  MOO settings. 
When all Service-Level Objectives (SLOs) cannot be satisfied, it guarantees max-min fairness over SLO satisfactions; otherwise, it degrades to WS for recommending a single solution.
Recent  optimization for cluster and cloud computing~\cite{LiNN14,ShiZLCLW14} 
focus on  running time of SQL queries, but not dataflow problems or multiple objectives.
Morpheus~\cite{JyothiCMNTYMGKK16} addresses the tradeoff between  cluster utilization and job's performance predictability by  codifying implicit user expectations as explicit SLOs and  enforces SLOs using scheduling techniques. 

WiseDB~\cite{MarcusP16,MarcusSP17} proposes learning-based techniques for cloud resource management.  A decision tree is trained on a set of performance and cost-related features collected from minimum cost schedules of sample workloads. 
Such minimum cost schedules are not available in our case. 
Dhalion~\cite{Floratou:2017:DSS} uses learning methods to detect backpressure and resolve it by tuning the degree of parallelism, but does not consider optimization for user objectives.
Li~\cite{Li:2018:MCD} minimizes end-to-end tuple processing time using deep reinforcement learning, and requires defining scheduling actions and the associated reward, which is not available in our problem.

More relevant is recent work on performance tuning and optimization.
PerfOrator~\cite{RajanKCK16}  and Ernest~\cite{Venkataraman:2016:EEP}  are modeling tools that use hand-crafted models while OtterTune~\cite{VanAken:2017:ADM,ZhangAWDJLSPG18}, which we leverage, learns more flexible models from the data. 
For optimizing a single objective, OtterTune builds a predictive model for each user query by mapping it to the most similar past query and based on the model, 
runs Gaussian Process exploration to minimize the objective. 
\cut{The newer work~\cite{MaAHMPG18} addresses prediction of the expected arrival rate of queries in the future based on historical data, hence orthogonal to our problem. }
CDBTune~\cite{Zhang:2019:EAC} solves a similar problem, but uses Deep RL to learn predictive models and determine the best configuration. 
\noindent
\textbf{Learning-based query optimization.}
Cardlearner~\cite{Wu:2018:TLO} uses a ML approach to learn cardinality models from previous job executions and uses them to predict the cardinalities in future jobs. 
Recent work~\cite{Marcus:2019:PDN} has used  neural networks  to match the structure of any optimizer-selected query execution plan and predicts latency.
Neo~\cite{Marcus:2019:NLQ} is a  DNN-based query optimizer 
that bootstraps its  optimization model from existing optimizers and then learns from incoming queries.

%% file: conclusions.tex
\section{Conclusions}

We presented a Progressive Frontier-based multi-objective optimizer that constructs a Pareto-optimal set of job configurations for multiple task-specific objectives, and recommends new job configurations to best meet them.
Using  batch and  streaming  workloads,  we showed that our MOO method outperform existing MOO methods~\cite{marler2004survey,Emmerich:2018:TMO} in both speed and coverage of the Pareto set, and outperforms  Ottertune~\cite{VanAken:2017:ADM}, a state-of-the-art approach, by a 26\%-49\% reduction in running time of the TPCx-BB benchmark, while  adapting to different application preferences on multiple objectives.
In future work, we plan to extend our optimizer to consider a pipeline of analytical tasks,  and optimize for both task-specific and system-wide objectives such as utilization and  throughput. 



%% file: ms.bbl
\begin{thebibliography}{10}

\bibitem{EC2}
Amazon ec2 instance types.
\newblock \url{ https://aws.amazon.com/ec2/instance-types/ }, 2019.

\bibitem{Aurora-serverless}
Amazon aurora serverless.
\newblock \url{https://aws.amazon.com/rds/aurora/serverless/}.

\bibitem{Azure}
Virtual machine sizes in azure.
\newblock \url{
  https://docs.microsoft.com/en-us/azure/virtual-machines/windows/sizes}, 2019.

\bibitem{Bonmin}
Bomin: Basic open-source nonlinear mixed integer programming.
\newblock \url{https://www.coin-or.org/Bonmin/}.

\bibitem{bussieck2003minlp}
M.~R. Bussieck and A.~Pruessner.
\newblock Mixed-integer nonlinear programming.
\newblock {\em SIAG/OPT Newsletter: Views \& News}, 14(1):19--22, 2003.

\bibitem{Couenne}
Couenne: Convex over and under envelopes for nonlinear estimation.
\newblock \url{https://projects.coin-or.org/Couenne/}.

\bibitem{Cplex}
Cplex optimizer.
\newblock \url{https://www.ibm.com/analytics/cplex-optimizer}.

\bibitem{das1997a}
I.~Das and J.~E. Dennis.
\newblock A closer look at drawbacks of minimizing weighted sums of objectives
  for pareto set generation in multicriteria optimization problems.
\newblock {\em Structural Optimization}, 14(1):63--69, 1997.

\bibitem{Deb:2002:FEM}
K.~Deb, A.~Pratap, S.~Agarwal, and T.~Meyarivan.
\newblock A fast and elitist multiobjective genetic algorithm: Nsga-ii.
\newblock {\em Trans. Evol. Comp}, 6(2):182--197, Apr. 2002.

\bibitem{DeWittG92}
D.~DeWitt and J.~Gray.
\newblock Parallel database systems: the future of high performance database
  systems.
\newblock {\em Commun. ACM}, 35(6):85--98, 1992.

\bibitem{DuanTB09}
S.~Duan, V.~Thummala, and S.~Babu.
\newblock Tuning database configuration parameters with ituned.
\newblock {\em {PVLDB}}, 2(1):1246--1257, 2009.

\bibitem{Flink-recovery2015}
S.~Dudoladov, C.~Xu, S.~Schelter, A.~Katsifodimos, S.~Ewen, K.~Tzoumas, and
  V.~Markl.
\newblock Optimistic recovery for iterative dataflows in action.
\newblock In {\em Proceedings of the 2015 {ACM} {SIGMOD} International
  Conference on Management of Data, Melbourne, Victoria, Australia, May 31 -
  June 4, 2015}, pages 1439--1443, 2015.

\bibitem{Emmerich:2018:TMO}
M.~T. Emmerich and A.~H. Deutz.
\newblock A tutorial on multiobjective optimization: Fundamentals and
  evolutionary methods.
\newblock {\em Natural Computing: an international journal}, 17(3):585--609,
  Sept. 2018.

\bibitem{Floratou:2017:DSS}
A.~Floratou, A.~Agrawal, B.~Graham, S.~Rao, and K.~Ramasamy.
\newblock Dhalion: Self-regulating stream processing in heron.
\newblock {\em Proc. VLDB Endow.}, 10(12):1825--1836, Aug. 2017.

\bibitem{gal16}
Y.~Gal and Z.~Ghahramani.
\newblock Dropout as a bayesian approximation: Representing model uncertainty
  in deep learning.
\newblock In M.~F. Balcan and K.~Q. Weinberger, editors, {\em Proceedings of
  The 33rd International Conference on Machine Learning}, volume~48 of {\em
  Proceedings of Machine Learning Research}, pages 1050--1059, New York, New
  York, USA, 20--22 Jun 2016. PMLR.

\bibitem{Ganguly:1992:QOP}
S.~Ganguly, W.~Hasan, and R.~Krishnamurthy.
\newblock Query optimization for parallel execution.
\newblock In {\em Proceedings of the 1992 ACM SIGMOD International Conference
  on Management of Data}, SIGMOD '92, pages 9--18, New York, NY, USA, 1992.
  ACM.

\bibitem{Liberti2018}
L.~iberti.
\newblock Undecidability and hardness in mixed-integer nonlinear programming.
\newblock Technical report, CNRS, Ecole Polytechnique, 2018.
\newblock \url{https://www.lix.polytechnique.fr/~liberti/rairo18.pdf}.

\bibitem{JMetal}
Jmetal: an object-oriented java-based framework for multi-objective
  optimization with metaheuristics.
\newblock \url{http://jmetal.sourceforge.net/}.

\bibitem{JyothiCMNTYMGKK16}
S.~A. Jyothi, C.~Curino, I.~Menache, S.~M. Narayanamurthy, A.~Tumanov,
  J.~Yaniv, R.~Mavlyutov, I.~Goiri, S.~Krishnan, J.~Kulkarni, and S.~Rao.
\newblock Morpheus: Towards automated slos for enterprise clusters.
\newblock In {\em 12th {USENIX} Symposium on Operating Systems Design and
  Implementation, {OSDI} 2016, Savannah, GA, USA, November 2-4, 2016.}, pages
  117--134, 2016.

\bibitem{kingma2014adam}
D.~P. Kingma and J.~Ba.
\newblock Adam: A method for stochastic optimization, 2014.

\bibitem{Kllapi:2011:SOD}
H.~Kllapi, E.~Sitaridi, M.~M. Tsangaris, and Y.~Ioannidis.
\newblock Schedule optimization for data processing flows on the cloud.
\newblock In {\em Proceedings of the 2011 ACM SIGMOD International Conference
  on Management of Data}, SIGMOD '11, pages 289--300, New York, NY, USA, 2011.
  ACM.

\bibitem{Knitro}
Artelys knitro user's manual.
\newblock \url{https://www.artelys.com/docs/knitro/index.html}.

\bibitem{LiDS15}
B.~Li, Y.~Diao, and P.~J. Shenoy.
\newblock Supporting scalable analytics with latency constraints.
\newblock {\em {PVLDB}}, 8(11):1166--1177, 2015.

\bibitem{LiNN14}
J.~Li, J.~F. Naughton, and R.~V. Nehme.
\newblock Resource bricolage for parallel database systems.
\newblock {\em {PVLDB}}, 8(1):25--36, 2014.

\bibitem{Li:2018:MCD}
T.~Li, Z.~Xu, J.~Tang, and Y.~Wang.
\newblock Model-free control for distributed stream data processing using deep
  reinforcement learning.
\newblock {\em Proc. VLDB Endow.}, 11(6):705--718, Feb. 2018.

\bibitem{Marcus:2019:NLQ}
R.~Marcus, P.~Negi, H.~Mao, C.~Zhang, M.~Alizadeh, T.~Kraska, O.~Papaemmanouil,
  and N.~Tatbul.
\newblock Neo: A learned query optimizer.
\newblock {\em Proc. VLDB Endow.}, 12(11):1705--1718, July 2019.

\bibitem{MarcusP16}
R.~Marcus and O.~Papaemmanouil.
\newblock Wisedb: {A} learning-based workload management advisor for cloud
  databases.
\newblock {\em {PVLDB}}, 9(10):780--791, 2016.

\bibitem{Marcus:2019:PDN}
R.~Marcus and O.~Papaemmanouil.
\newblock Plan-structured deep neural network models for query performance
  prediction.
\newblock {\em Proc. VLDB Endow.}, 12(11):1733--1746, July 2019.

\bibitem{MarcusSP17}
R.~Marcus, S.~Semenova, and O.~Papaemmanouil.
\newblock A learning-based service for cost and performance management of cloud
  databases.
\newblock In {\em 33rd {IEEE} International Conference on Data Engineering,
  {ICDE} 2017, San Diego, CA, USA, April 19-22, 2017}, pages 1361--1362, 2017.

\bibitem{marler2004survey}
R.~Marler and J.~S. Arora.
\newblock Survey of multi-objective optimization methods for engineering.
\newblock {\em Structural and Multidisciplinary Optimization}, 26(6):369--395,
  2004.

\bibitem{messac2012from}
A.~Messac.
\newblock From dubious construction of objective functions to the application
  of physical programming.
\newblock {\em AIAA Journal}, 38(1):155--163, 2012.

\bibitem{messac2003nc}
A.~Messac, A.~Ismailyahaya, and C.~A. Mattson.
\newblock The normalized normal constraint method for generating the pareto
  frontier.
\newblock {\em Structural and Multidisciplinary Optimization}, 25(2):86--98,
  2003.

\bibitem{NEOS-guide}
Neos guide: Nonlinear programming software.
\newblock \url{https://neos-guide.org/content/nonlinear-programming}.

\bibitem{NEOS-server}
Neos solvers: Nonlinear programming software.
\newblock \url{https://neos-server.org/neos/solvers/index.html}.

\bibitem{RajanKCK16}
K.~Rajan, D.~Kakadia, C.~Curino, and S.~Krishnan.
\newblock Perforator: eloquent performance models for resource optimization.
\newblock In {\em Proceedings of the Seventh {ACM} Symposium on Cloud
  Computing, Santa Clara, CA, USA, October 5-7, 2016}, pages 415--427, 2016.

\bibitem{SchulzSK2018}
E.~Schulz, M.~Speekenbrink, and A.~Krause.
\newblock A tutorial on gaussian process regression: Modelling, exploring, and
  exploiting functions.
\newblock {\em Journal of Mathematical Psychology}, 85:1--16, August 2018.

\bibitem{ShiZLCLW14}
J.~Shi, J.~Zou, J.~Lu, Z.~Cao, S.~Li, and C.~Wang.
\newblock Mrtuner: {A} toolkit to enable holistic optimization for mapreduce
  jobs.
\newblock {\em {PVLDB}}, 7(13):1319--1330, 2014.

\bibitem{Song2020}
F.~Song, K.~Zaouk, C.~Lyu, A.~Sinha, Q.~Fan, Y.~Diao, and P.~Shenoy.
\newblock Boosting cloud data analytics using multi-objective optimization.
\newblock Technical report, 2020.
\newblock \url{https://hal.inria.fr/hal-02549758}.

\bibitem{tan2016tempo}
Z.~Tan and S.~Babu.
\newblock Tempo: robust and self-tuning resource management in multi-tenant
  parallel databases.
\newblock {\em Proceedings of the VLDB Endowment}, 9(10):720--731, 2016.

\bibitem{TPCx-BB}
TPCx-BB.
\newblock Tpcx-bb (bigbench) benchmark for big data analytics.
\newblock \url{http://www.tpc.org/tpcx-bb/}.

\bibitem{Trummer:2014:ASM}
I.~Trummer and C.~Koch.
\newblock Approximation schemes for many-objective query optimization.
\newblock In {\em Proceedings of the 2014 ACM SIGMOD International Conference
  on Management of Data}, SIGMOD '14, pages 1299--1310, New York, NY, USA,
  2014. ACM.

\bibitem{TrummerK15}
I.~Trummer and C.~Koch.
\newblock An incremental anytime algorithm for multi-objective query
  optimization.
\newblock In {\em Proceedings of the 2015 {ACM} {SIGMOD} International
  Conference on Management of Data, Melbourne, Victoria, Australia, May 31 -
  June 4, 2015}, pages 1941--1953, 2015.

\bibitem{VanAken:2017:ADM}
D.~Van~Aken, A.~Pavlo, G.~J. Gordon, and B.~Zhang.
\newblock Automatic database management system tuning through large-scale
  machine learning.
\newblock In {\em Proceedings of the 2017 ACM International Conference on
  Management of Data}, SIGMOD '17, pages 1009--1024, New York, NY, USA, 2017.
  ACM.

\bibitem{Venkataraman:2016:EEP}
S.~Venkataraman, Z.~Yang, M.~Franklin, B.~Recht, and I.~Stoica.
\newblock Ernest: Efficient performance prediction for large-scale advanced
  analytics.
\newblock In {\em Proceedings of the 13th Usenix Conference on Networked
  Systems Design and Implementation}, NSDI'16, pages 363--378, Berkeley, CA,
  USA, 2016. USENIX Association.

\bibitem{Wu:2018:TLO}
C.~Wu, A.~Jindal, S.~Amizadeh, H.~Patel, W.~Le, S.~Qiao, and S.~Rao.
\newblock Towards a learning optimizer for shared clouds.
\newblock {\em Proc. VLDB Endow.}, 12(3):210--222, Nov. 2018.

\bibitem{Zaharia+2012:rdd}
M.~Zaharia, M.~Chowdhury, T.~Das, A.~Dave, J.~Ma, M.~McCauley, M.~J. Franklin,
  S.~Shenker, and I.~Stoica.
\newblock Resilient distributed datasets: a fault-tolerant abstraction for
  in-memory cluster computing.
\newblock In {\em Proceedings of the 9th USENIX conference on Networked Systems
  Design and Implementation}, NSDI'12, pages 2--2, Berkeley, CA, USA, 2012.
  USENIX Association.

\bibitem{Zaharia+2013:discretized}
M.~Zaharia, T.~Das, H.~Li, T.~Hunter, S.~Shenker, and I.~Stoica.
\newblock Discretized streams: fault-tolerant streaming computation at scale.
\newblock In {\em Proceedings of the Twenty-Fourth ACM Symposium on Operating
  Systems Principles}, SOSP '13, pages 423--438, New York, NY, USA, 2013. ACM.

\bibitem{UDAO-vldb-demo}
K.~Zaouk, F.~Song, C.~Lyu, A.~Sinha, Y.~Diao, and P.~J. Shenoy.
\newblock {UDAO:} {A} next-generation unified data analytics optimizer (vldb
  2019 demo).
\newblock {\em {PVLDB}}, 12(12):1934--1937, 2019.

\bibitem{ZhangAWDJLSPG18}
B.~Zhang, D.~V. Aken, J.~Wang, T.~Dai, S.~Jiang, J.~Lao, S.~Sheng, A.~Pavlo,
  and G.~J. Gordon.
\newblock A demonstration of the ottertune automatic database management system
  tuning service.
\newblock {\em {PVLDB}}, 11(12):1910--1913, 2018.

\bibitem{Zhang:2019:EAC}
J.~Zhang, Y.~Liu, K.~Zhou, G.~Li, Z.~Xiao, B.~Cheng, J.~Xing, Y.~Wang,
  T.~Cheng, L.~Liu, M.~Ran, and Z.~Li.
\newblock An end-to-end automatic cloud database tuning system using deep
  reinforcement learning.
\newblock In {\em Proceedings of the 2019 International Conference on
  Management of Data}, SIGMOD '19, pages 415--432, New York, NY, USA, 2019.
  ACM.

\end{thebibliography}
